\documentclass[%
 reprint,
 amsmath,amssymb,
 aps,
]{revtex4-1}

\usepackage{ulem}
\usepackage[Symbol]{upgreek}
\usepackage{amsmath}
\usepackage{amssymb}%花体字母加粗
\usepackage{mathrsfs}%花体字母
\usepackage{graphicx}% Include figure files
\usepackage{dcolumn}% Align table columns on decimal point
\usepackage{bm}% bold math
\usepackage{hyperref}% add hypertext capabilities
\newcommand{\ket}[1]{\ensuremath{\left|#1\right\rangle}}

\usepackage{xcolor}

\newcommand{\beginsupplement}{%
        \setcounter{table}{0}
        \renewcommand{\thetable}{S\arabic{table}}%
        \setcounter{figure}{0}
        \renewcommand{\thefigure}{S\arabic{figure}}%
     }

\begin{document}

\preprint{APS/123-QED}

\title{High-dimensional classically entangled light from a laser}% Force line breaks with \\
%\thanks{Subtitle: Towards high-dimensional classical entanglement.}%

\author{Yijie Shen$^{1,2,3}$, Isaac Nape$^{1}$, Xilin Yang$^{4}$, Xing Fu$^{3,5}$, Mali Gong$^{3,5}$, Darryl Naidoo$^{1,6}$, and Andrew Forbes$^{1}$}
\affiliation{
	{$^{1}$School of Physics, University of the Witwatersrand, Private Bag 3, Wits 2050, South Africa}\\{$^{2}$State Key Laboratory of Precision Measurement Technology and Instruments, Department of Precision Instrument, Tsinghua University, Beijing 100084, China}\\{$^{3}$Optoelectronics Research Centre, University of Southampton, Southampton SO17 1BJ, United Kingdom}\\{$^{4}$Electrical and Computer Engineering Department, University of California, Los Angeles, CA 90095, USA}\\{$^{5}$Key Laboratory of Photonic Control Technology (Tsinghua University), Ministry of Education, Beijing 100084, China}\\
	{$^{6}$CSIR National Laser Centre, PO Box 395, Pretoria 0001, South Africa}
	%{$^\ast$Corresponding author: andrew.forbes@wits.ac.za}\\
}

\date{\today}

\begin{abstract}
\noindent \textbf{Vectorially structured light has emerged as an enabling tool in many diverse applications, from communication to imaging, exploiting quantum-like correlations courtesy of a non-separable spatially varying polarization structure.  Creating these states at the source remains challenging and is presently limited to two-dimensional vectorial states by customized lasers.   Here we invoke ray-wave duality in a simple laser cavity to produce polarization marked multi-path modes that are non-separable in three degrees of freedom and in eight dimensions.  As a topical example, we use our laser to produce the complete set of Greenberger-Horne-Zeilinger (GHZ) basis states, mimicking high-dimensional multi-partite entanglement with classical light, which we confirm by a new projection approach. We offer a complete theoretical framework for our laser based on SU(2) symmetry groups, revealing a rich parameter space for further exploitation. Our approach requires only a conventional laser with no special optical elements, is easily scaleable to higher dimensions, and offers a simple but elegant solution for at-the-source creation of classically entangled states of structured light, opening new applications in simulating and enhancing high-dimensional quantum systems.}

\end{abstract}

\maketitle

%\tableofcontents
\noindent A quintessential property of quantum systems is the notion of entanglement between particles, namely, that the measurement of some property of one particle effects the measurement outcome of another.  This has been demonstrated with many systems, notably photonic states of light.  The hallmark of entangled states is non-separability, but  non-separability is not unique to quantum mechanics: it is found elsewhere in nature, and in particular it may be engineered in the form of vectorial states of structured light, controversially referred to as classically entangled light~\cite{spreeuw1998classical,forbes2019classically,konrad2019quantum,toninelli2019concepts}.  Here the entanglement is local, between internal degrees of freedom (DoFs) of one classical light field,  and so cannot capture the non-locality property of quantum entangled particles.  Yet the notion of classical entanglement is more than simple mathematical machinery, with such states of light having violated Bell-like inequalities \cite{kagalwala2013bell}, used for classical teleportation \cite{Silva2016}, enabled quantum walks with bright light \cite{sephton2019versatile,d2020two}, used to realize real-time quantum error correction in noisy channels \cite{ndagano2017characterizing}, and exploited for their classical vectorial properties in a myriad of applications \cite{rosales2018review}, including metrology~\cite{Toppel2014,d2013photonic}, spin-to–orbital conversion in metamaterial~\cite{devlin2017arbitrary}, multi-channel high-capacity communication~\cite{liu2018direct,willner2018vector,wangcomplete},  trapping~\cite{skelton2013trapping,min2013focused}, super-resolution imaging~\cite{kozawa2018superresolution}, and kinematic sensing~\cite{berg2015classically}.  

Given the many applications that such classically entangled states have fostered, much attention has gone into their creation.  A myriad of tools exist for creation external to the source \cite{zhan2009cylindrical,rubano2019q,wang2018recent}, but with only limited success directly from lasers \cite{forbes2019structured}, with demonstrations limited to two-dimensional (qubit equivalent) non-separable states by special intra-cavity elements in the form of liquid crystal geometric phase elements \cite{naidoo2016controlled}, metasurfaces \cite{maguid2018topologically} as well as the use of optical fibre \cite{mao2017ultrafast} and on-chip solutions that exploit topological \cite{cai2012integrated,zambon2019optically,miao2016orbital} or organic \cite{stellinga2018organic} structures. In contrast, the ability to access arbitrarily engineered high-dimensional state spaces with vectorial light would be highly beneficial, opening the way to many exciting applications \cite{aiello2015quantum,mabena2017high,mabena2020quantum,korolkova2019quantum,forbes2019quantum,shen2019optical}.  

%All the aforementioned examples exploited polarization and spatial mode as the two DoFs, so-called vector beams, often expressed in terms of spin ($\pm \sigma$) and orbital ($\pm \ell$) angular momentum components of light (vector vortex beams).  This is akin to a two-dimensional (2D) quantum entangled state because it has two component modes in two orthogonal bases, e.g., $ \ket{\ell}_A\ket{\sigma}_B + \ket{-\ell}_A\ket{-\sigma}_B $.  However, such vector beams are naturally limited to a 2-dimensional representation by virtue of the two polarization states, and can be described as states on a 2D space, e.g., a sphere \cite{hps1,hyps1}.  In contrast, the ability to access arbitrarily engineered high-dimensional state spaces with vectorial light would be highly beneficial, opening the way to many exciting applications \cite{aiello2015quantum}.  %Thus, breaking the inherent dimension limit of vector beams, and doing so in a compact solution directly from a laser, remains an open challenge. 

Here we invoke complementary ray-wave duality in a simple laser cavity to produce polarization marked multi-path modes for high-dimensional classically entangled structured light directly from the source.  In our cavity, the oscillating modes appear as geometric ray orbits (patterns) along the length and transverse to the cavity, each with a fixed shape and period determined by the cavity length and pump light position.  We introduce the concept of gain control to realize polarization marking of the ray orbits without the need for any internal polarizing elements.  Although each exiting ray appears to follow an independent path upon propagation, the output of all paths from the same geometric shape is a single transverse mode with spatial coherence, courtesy of the wave picture.  Thus we use the DoFs of cavity ray direction, location and polarization to realize eight dimensional classically entangled light from a laser, which we go on to tailor in amplitude and phase, demonstrating the first complete set of classically entangled Greenberger-Horne-Zeilinger (GHZ) states \cite{greenberger1990bell}.  Our work overcomes the two-dimensional (2D) limitation of existing laser solutions, creates new states of non-separable structured light not proposed before, and does so in an otherwise empty laser cavity without any special elements.  The birth of high-dimensional classical entanglement states from such a simple and compact laser paves the way for exciting new applications of vectorially structured light.
\begin{figure*}[t!]
	\centering
	\includegraphics[width=0.9\linewidth]{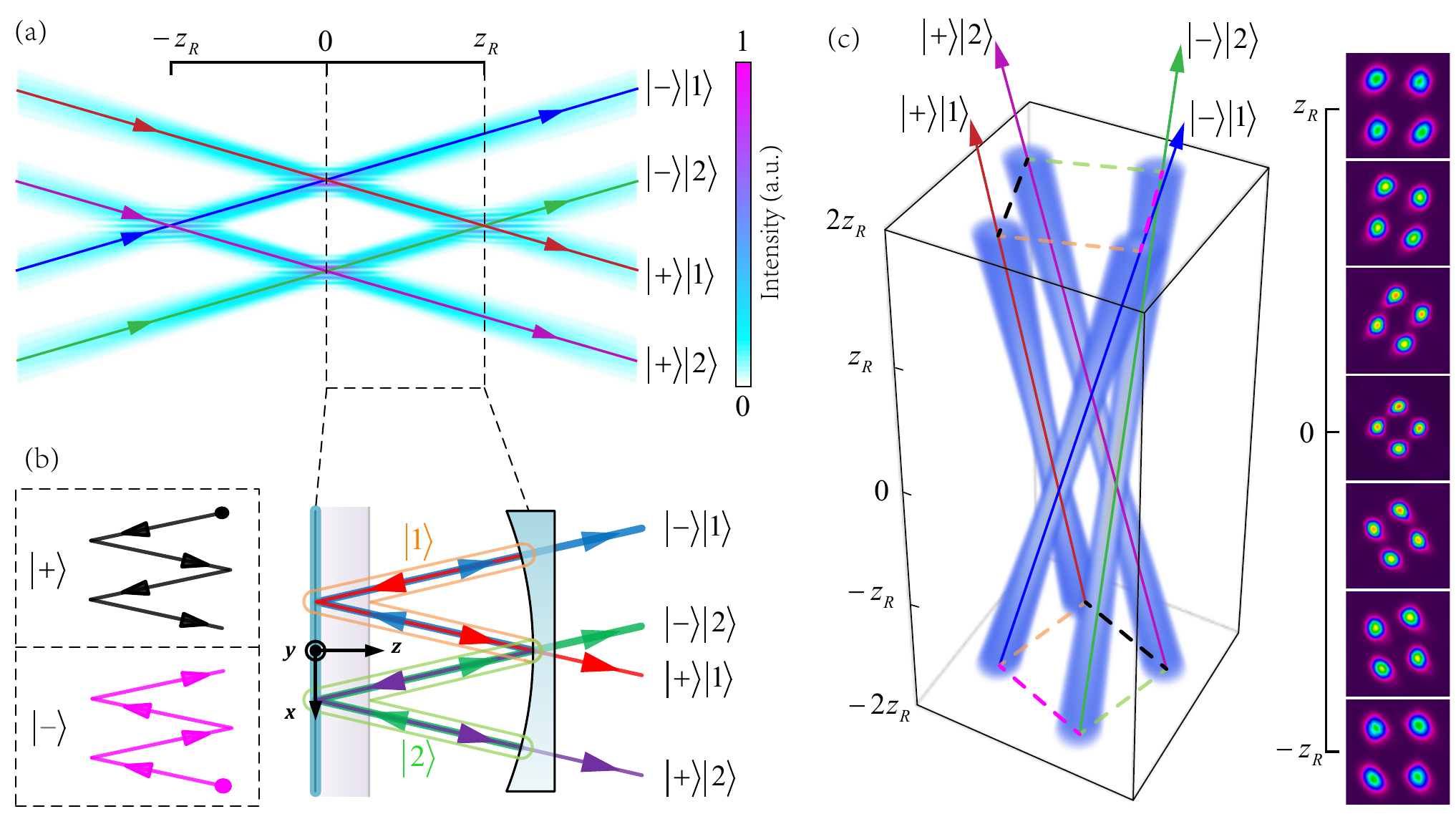}
	\caption{\textbf{Laser concept.}  (a) A 2D planar representation of the desired geometric mode, where the mode evolves from wave-like fringes (at $z=0$ and $\pm z_R$) to a ray-like trajectory.  The means to create this mode in the cavity is illustrated in (b), where completed oscillation orbits can be described by the direction states $\ket{+}$ and $\ket{-}$ (equivalently sub-orbits shown in black and pink, respectively, in the dash-line boxes), as well as paths originating from ray positions $\ket{1}$ and $\ket{2}$ (equivalently sub-orbits shown in orange and green, respectively). (c) A 3D schematic of a skewed (vortex) SU(2) geometric mode propagating in free space, with experimental beam images shown at example propagation distances.  The planar mode is converted to the skewed version with a pair of cylindrical lenses.}
	\label{concept}\label{f1}
\end{figure*}

\section*{Results}
\noindent \textbf{Concept.}  Spatial mode control in lasers allows one to specify the desired modal properties of amplitude, phase and polarization.  This may be as vector or scalar states of light, but pertinently, there are only two DoFs with which to execute the control: the transverse mode and its polarization, conveniently mapped to a sphere \cite{hps1,hyps1}.  As such, only analogues to qubits have been demonstrated from lasers \cite{forbes2019structured}.  The challenge is to find more DoFs in a system seemingly constrained to just two, a requirement in order to reach higher dimensions in classically entangled light, but paradoxically, we wish to do so in an otherwise empty laser configuration with no special elements.  We will show that we can select our new DoFs to be the oscillating direction, round trip location, and polarization so that we can mimic multi-partite high-dimensional quantum states, classically.  Our concept utilizes the ability to produce an SU(2) coherent state from the laser, treating the cavity as a quantum harmonic oscillator with both wave-packet and classical ray trajectory interpretations \cite{su22,su23,rwd3}.  We do so by engineering a frequency-degenerate plano-concave cavity whose output mode can be shown to be exactly the desired coherent state (see Supplementary Material).  Importantly, when the resonator operates in this frequency-degenerate state and the gain is excited off-axis (at some localized position) the laser mode too becomes localized on periodic trajectories that are controlled by the off-axis pump position.  While the output is a coherent state mode and is wave-like, the ray trajectories define particle-like positions of localized intensity.  This ray-wave duality allows us to label the output in terms of the ray trajectories rather than the modes themselves.

%\Note{Moreover, the phase and intensity on the different SU(2) orbits can also be modulated. After arbitrary polarization, phase, and intensity modulations on various orbits, a normal SU(2) geometric beam would be transformed into a complex vector structured beam. We call the modulated beam as \textit{general SU(2) vector beam}, which can no longer be described by a simple SU(2) coherent state but a complex 8-D state with the bases of $|-\rangle|1\rangle|D\rangle$, $|-\rangle|1\rangle|A\rangle$, $|-\rangle|2\rangle|D\rangle$, $|-\rangle|2\rangle|A\rangle$, $|+\rangle|1\rangle|D\rangle$, $|+\rangle|1\rangle|A\rangle$, $|+\rangle|2\rangle|D\rangle$, and $|+\rangle|2\rangle|A\rangle$, where the two eigenstates of polarization, diagonal and anti-diagonal polarization states ($|D\rangle$ and $|A\rangle$) can also be replaced by horizontal and vertical polarizations ($|H\rangle$ and $|V\rangle$) or right- and left-handed circular polarizations ($|R\rangle$ and $|L\rangle$).}
%When a laser cavity fulfils a special geometric condition where the longitudinal and transverse frequencies are coupled to each other (a frequency-degenerate state), the wave-packet of the laser mode can equally be described in a ray-like picture by a geometric oscillating trajectory (our geometric mode) with two DoFs, the oscillating direction and the round trip location (see the Supplementary Material for the degenerate orbits and their mathematical description).  

\begin{figure}
	\includegraphics[width=\linewidth]{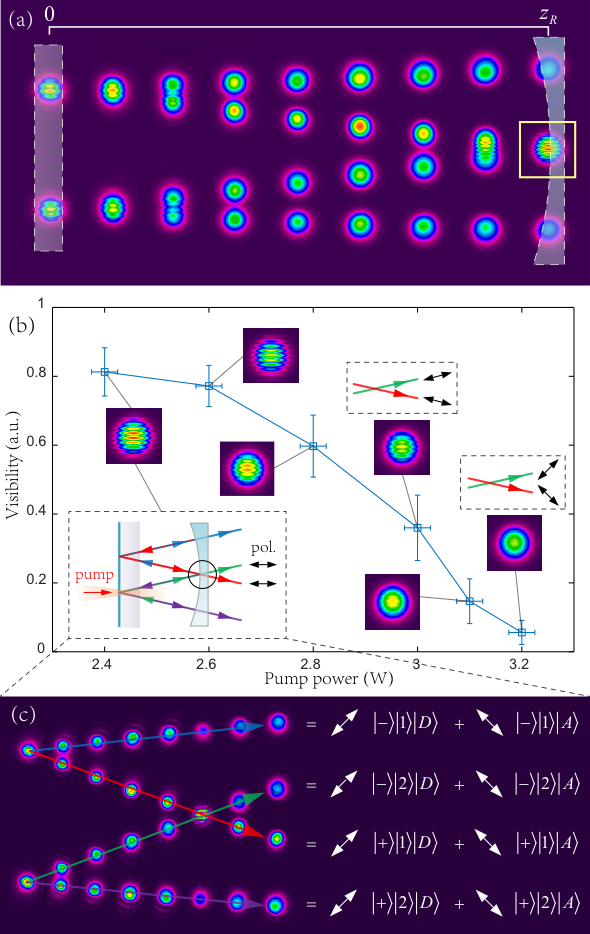}
	\caption{\textbf{Vectorial ray-wave beams.} (a) Experimental verification of an SU(2) geometric mode exhibiting ray-wave duality, where the corresponding positions of the flat and concave cavity mirrors are shown.  The images show the evolution of the output mode from one end of the cavity to the other as a sequence of camera images.  When the rays overlap, the wave-like nature is evident, as seen by the fringes shown in the yellow box. (b) The polarization structured of the rays can be adjusted by tuning the pumping, inducing a change in fringe visibility with the pump power as the polarization state of each ray is modified. The insets show the measured interference fringes at $z=z_R$ for selected pump powers. The dashed boxes depict the corresponding ray representation together with polarization change. (c) The notation of all the elements in eight dimensions in an output vectorial ray-wave beam.} 
	\label{fm2}
\end{figure}

For instance, one such trajectory is shown in Fig.~\ref{concept}, with the forward ($\ket{+}$) and backward ($\ket{-}$) oscillating states and the first ($\ket{1}$) and second ($\ket{2}$) roundtrip location state forming a completed oscillation with the potential to reach dimension $d = 4$ (see Supplementary Material for the other geometries to reach higher dimensions). In Fig.~\ref{concept} (a) we show a 2D simulation of this ray-like propagation with marked trajectories. At $z=0$ and $\pm z_R$ ($z_R$ is the Rayleigh range) the wave-like behaviour is evident by the fringes in the beam as trajectories overlap. To understand how to create this from a laser, we highlight the generation step inside a laser in Fig.~\ref{concept} (b).  Here the output four rays originating from two points ($\ket{1},\ket{2}$) on the rear mirror comprises two V-shaped locations, and the pair of rays in a certain location state has two different directions ($\ket{+},\ket{-}$), so that the output is spanned by the basis states ${\cal H}_{\text{ray}} \in \{ \ket{+}\ket{1},\ket{-}\ket{1},\ket{+}\ket{2},\ket{-}\ket{2} \}$, a 4-dimensional Hilbert space, while in the wave picture the emitted output from the laser is a single spatial mode.  The planar mode can be converted by simple cylindrical lenses to a skewed ray version shown in Fig.~\ref{concept} (c), with the physical implementation in the laboratory shown in the Supplementary Material, section~\ref{AJ}.  Alongside the simulation is the measured beam profile at selected propagation distances outside the cavity.  We can further verify the laser operation by imaging the internal cavity mode to planes external to the laser for ease of measurement, shown for a period-4 trajectory example (frequency-degenerate state $\ket{\Omega = 1/4}$) experimentally in Figure~\ref{fm2} (a). The results show the tell-tale signs of a wave-like behaviour, as evident by the interference fringes with high visibility at $z=0$ and at $z=z_R$, while appearing to be independent ray trajectories.  The output is a 4-dimensional scalar state.

Since one of the DoFs is polarization, our scheme requires us to engineer the cavity to have trajectory-dependent polarization control without any intra-cavity elements.  We achieve this by exploiting the fact that differing ray trajectories will impinge on the laser crystal at angles other than normal, as well as thermal effects as a function of pump power (see Supplementary Materials, section~\ref{AI}).  By deploying a c-cut crystal which exhibits angle dependent birefringence at non-normal incident angles, we can mark our orbits with polarization by simple adjustment of the pump light position and power, with results shown in Fig.~\ref{fm2} (b). The visibility of the fringes decreases to zero as the pump power is increased (for a particular angle of trajectory) as a result of orthogonal ray polarization states. The power control of the gain results in all rays evolving from an entirely linear horizontal state to a vector state of diagonal ($\ket{D}$) and anti-diagonal ($\ket{A}$) polarizations.  The results confirm that the cavity can be forced into a ray-like mode and further, that the various orbits can be individually modulated in polarization, resulting in a polarization structured SU(2) vector beam.  We derive a general analytical form for the output state [see Eq.~(\ref{gv}) in the Supplementary Material, section~\ref{AF}]; for the example geometry of Figs.~\ref{concept} and \ref{fm2} it is a laser mode in an 8-dimensional Hilbert space spanned by the basis states ${\cal H}_8 \in \{ \ket{+}\ket{1}\ket{D}, \ket{-}\ket{1}\ket{D}, \ket{+}\ket{2}\ket{D},$ $\ket{-}\ket{2}\ket{D}, \ket{+}\ket{1}\ket{A}, \ket{-}\ket{1}\ket{A}, \ket{+}\ket{2}\ket{A}, \ket{-}\ket{2}\ket{A} \}$.  We show this full state experimentally in Fig.~\ref{fm2} (c).

\vspace{0.5cm}

\begin{figure*}[t!]
	\centering
	\includegraphics[width=\linewidth]{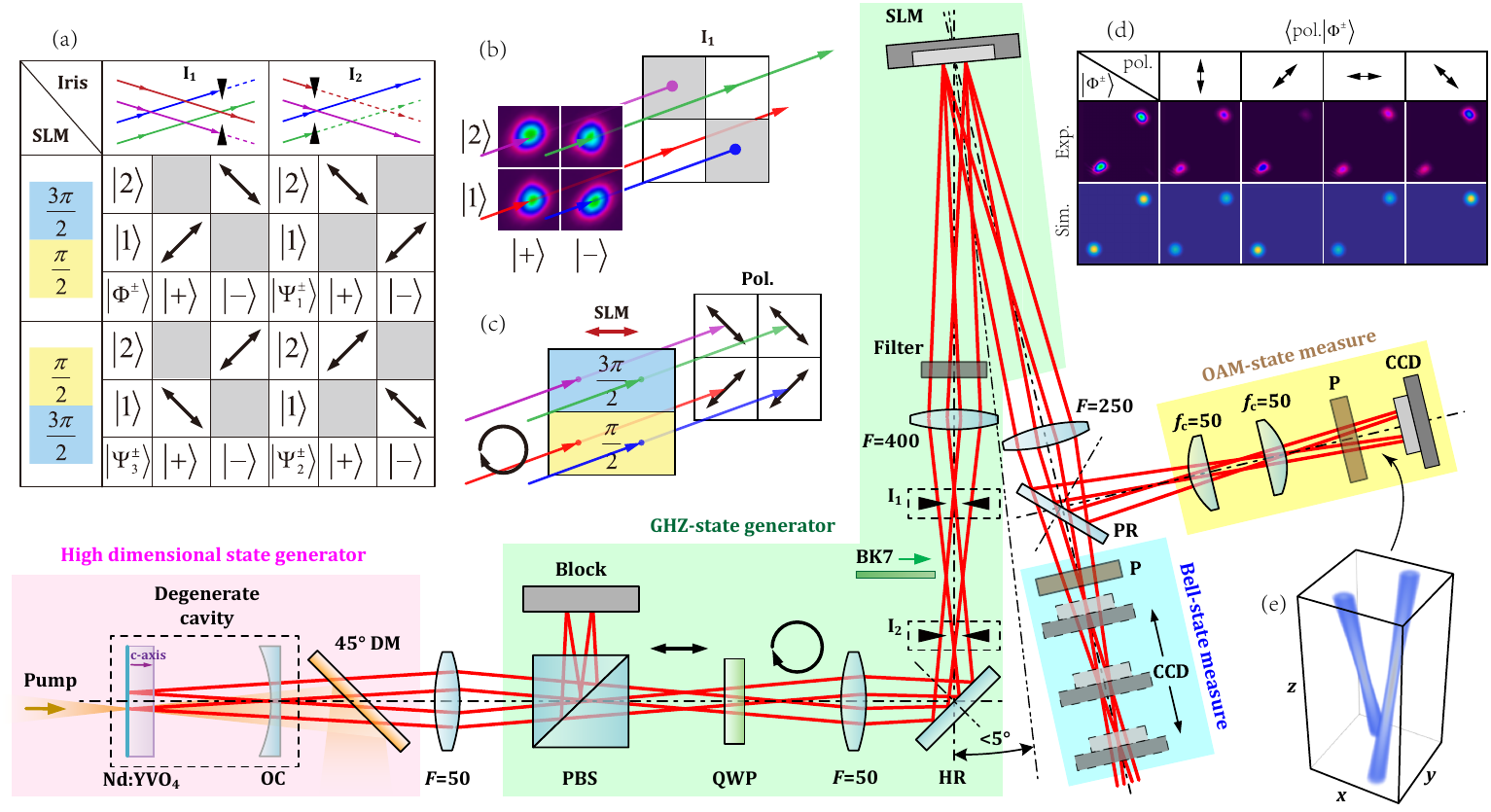}
	\caption{\textbf{Classical GHZ states.} The experimental setup used to generate classical GHZ states from our laser, including the core steps of high-dimensional state generation (laser), GHZ-state generation, and two measurement steps to confirm the state properties. (a) shows the required path and polarization transformations needed for each GHZ state, performed by an iris (located at I$_1$ or I$_2$) and SLM phases ($3\pi/2$ and $\pi/2$), respectively.  In (b) and (c) this is unpacked graphically: the incoming four lobes are reduced to two, and which are subsequently altered in polarization.  (d) shows results for the vector beam corresponding to the first maximally entangled group $|\Phi^\pm\rangle$ both experimentally (Exp.) and simulated (Sim.).  The arrows depict the orientation of the polarizer in the measurement stage of the OAM-state measure. In the tomography measurement (Bell-state measure), each of the eight GHZ states can be inferred by just a polarizer and a CCD camera. The CCD camera is moved to different locations and captures the interferometric fringes for a visibility calculation. (OC: output coupler mirror, DM: dichroic mirror, PBS: polarization splitting prism, QWP: quarter-wave plate, HR: high-reflective mirror, PR: partial-reflective mirror, SLM: spatial light modulator, CCD: charge-coupled device camera, P: Polarizer.)}
	\label{f3}
\end{figure*}
\begin{figure*}[t!]
	\centering
	\includegraphics[width=\linewidth]{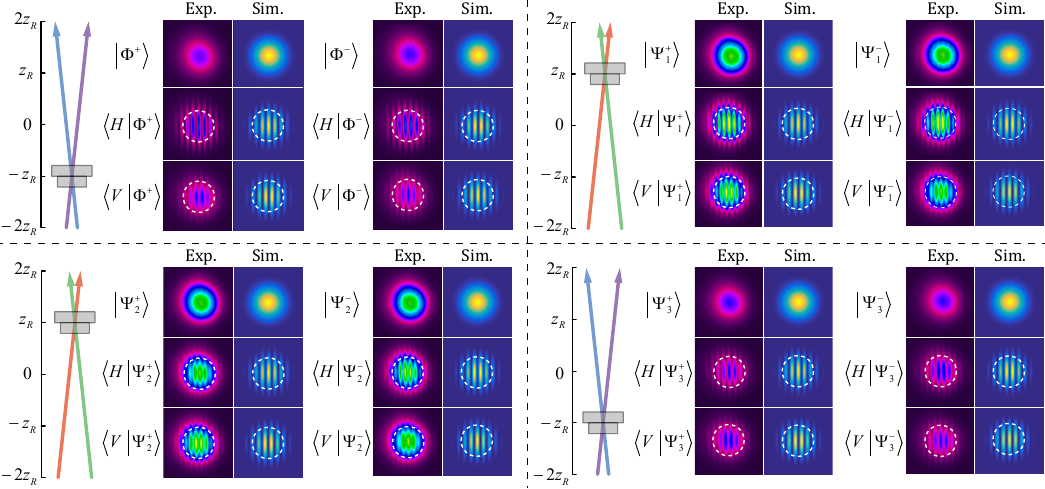}
	\caption{\footnotesize \textbf{Identification of each GHZ state.} The experimental (Exp.) and theoretical (Sim.) results showing interference patterns at the overlapped region of the two orbits corresponding to each GHZ state. The plot on the left for each group shows the location of the CCD camera for capturing the interference fringes. $|\Phi^+\rangle$: no fringe for the original state, center-bright fringes would be observed after projecting polarization on $|H\rangle$ state i.e. for  $\langle H|\Phi^+\rangle$, center-dark fringes for $\langle V|\Phi^+\rangle$; $|\Phi^-\rangle$: center-dark fringes for $\langle H|\Phi^-\rangle$, center-bright fringes for $\langle V|\Phi^-\rangle$;  $|\Psi_1^+\rangle$: center-bright fringes for $\langle H|\Psi_1^+\rangle$, center-dark fringes for $\langle V|\Psi_1^+\rangle$; $|\Psi_1^-\rangle$: center-dark fringes for $\langle H|\Psi_1^-\rangle$, center-bright fringes for $\langle V|\Psi_1^-\rangle$; $|\Psi_2^+\rangle$: center-bright fringes for $\langle H|\Psi_2^+\rangle$, center-dark fringes for $\langle V|\Psi_2^+\rangle$; $|\Psi_2^-\rangle$: center-dark fringes for $\langle H|\Psi_2^-\rangle$, center-bright fringes for $\langle V|\Psi_2^-\rangle$; $|\Psi_3^+\rangle$: center-bright fringes for $\langle H|\Psi_3^+\rangle$, center-dark fringes for $\langle V|\Psi_3^+\rangle$; $|\Psi_3^-\rangle$: center-dark fringes for $\langle H|\Psi_3^-\rangle$, center-bright fringes for $\langle V|\Psi_3^-\rangle$.}
	\label{Bell}
\end{figure*}

\noindent \textbf{GHZ states from a laser.} The results shown thus far confirm that this simple laser cavity (with no special internal elements) is able to lase on an SU(2) vector beam expressed in an 8-dimensional Hilbert space with three DoFs. In our example results, this is achieved through oscillation on the frequency-degenerate state $\ket{\Omega = 1/4}$ and phase state $|\phi = \pi\rangle$ but higher dimensions are possible by small adjustments to the cavity length and pump position (see Supplementary Material).  Now we show that it is possible to controllably modulate this high-dimensional laser state to some desired state; by way of example we create the first classically entangled GHZ states, with the concept and setup shown in Fig.~\ref{f3}.  In particular we wish to create a complete set of GHZ states, namely, $\ket{\Phi^\pm}$, $\ket{\Psi^\pm_1}$, $\ket{\Psi^\pm_2}$, and $\ket{\Psi^\pm_3}$, with full details given in the Supplementary Material, section~\ref{AG}.

The idea is to externally modulate the output state from the laser to engineer the amplitude and phase of each term independently, producing the general state 
\begin{eqnarray}
\nonumber
\ket{u_8} &=& \alpha_1 \ket{+}\ket{1}\ket{D} + \alpha_2 \ket{-}\ket{1}\ket{D} + \alpha_3 \ket{+}\ket{2}\ket{D} \\ \nonumber
&&  +  \alpha_4 \ket{+}\ket{1}\ket{A}  + \alpha_5 \ket{-}\ket{2}\ket{D} + \alpha_6 \ket{-}\ket{1}\ket{A} \\ \nonumber
 && + \alpha_7 \ket{+}\ket{2}\ket{A} + \alpha_8 \ket{-}\ket{2}\ket{A} ,  \nonumber
 \label{eq:generalState}
\end{eqnarray}
\noindent and converting it to each of the eight GHZ basis states.  For example, the transformation
\begin{equation}
\ket{u_8} \rightarrow \ket{\Phi^\pm} =  \frac{\ket{+}\ket{1}\ket{D} \pm \ket{-}\ket{2}\ket{A}}{\sqrt{2}} \nonumber
\label{eq-GHZ1}
\end{equation} 
\noindent requires a modulation that sets all amplitudes to zero except $|\alpha_1| = |\alpha_8|=\frac{1}{\sqrt{2}}$ and a relative phase shift between the two decomposed ray modes.  The general setup to achieve this (and other modulations) is shown in Fig.~\ref{f3}.  The main experimental arrangement includes the laser for creating the initial high-dimensional state followed by a tailoring step to convert it into specific desirable classes, here the GHZ states are generated as an example.  Finally, the states are directed to two measurement devices: the vectorial nature of the prepared states are measured by a polarizer and camera (OAM-state measure) and a Bell-state measurement device, which we introduce here, is used for the tomographic projections. The four kinds of vector beams corresponding to the four maximum entangled groups of the eight GHZ states are illustrated by the graphical procedure as Fig.~\ref{f3}~(a), and unpacked in parts (b) and (c).  Here we outline the process using the $\ket{\Phi^\pm}$ state as an example, with full details for all cases given in the Methods and Supplementary Material.   We switch from a linear polarization basis to circular with a quarter-wave plate (QWP), eliminate light on the ray states $\ket{+}\ket{2}$ and $\ket{-}\ket{1}$ by iris I$_1$ and then modulate the polarization of the $\ket{+}\ket{1}$ and $\ket{-}\ket{2}$ paths into diagonal and anti-diagonal states using programmed phases on a spatial light modulator. Similar transformations allow us to generate all GHZ states in the complete family of maximally entangled states. The controllable generation of all the GHZ states provides further verification that the general SU(2) beam in our system is indeed expressed in an 8-dimensional space, as GHZ states form a complete basis in 8-dimensions.

The experimental images of skewed rays, together with the theoretical predictions, are shown in Fig.~\ref{f3}~(d) for the $\ket{\Phi^\pm}$ states as a function of the orientation of the polarizer in the OAM-state measure step, showing excellent agreement. The intensity distribution is a two-lobed structure, consistent with the corresponding GHZ state, while the evolution of the lobe intensities confirm the vectorial nature of the field.  A reconstruction of the GHZ-state mode's propagation in free-space is shown in Fig.~\ref{f3}~(e).  Results for all other states are shown in the Supplementary Material, Fig.~\ref{P}.    

%The experimental and simulated results of this state modulation is shown in Figure~\ref{f3} (a) for the $\ket{\Phi^\pm}$ state.  The intensity distribution is a two-lobed structure, consistent with the GHZ state. \YS{All the possible polarization distributions obtained in our system are listed in the table as Figure~\ref{f3}(b). To measure the vectorial fields of the GHZ-state vector beams, we used an OAM convertor to introduce the multi-lobe structures and a polarizer to measure the polarization distributions, shown in Figure~\ref{f3}(c).} The system involves first passing the beam through an astigmatic mode converter to convert the planar SU(2)-like beam into a vortex beam with disjoint geometric orbits followed by a polarization projection (see Methods). A combination of these measurements comprise a tomography for our vector beams corresponding to the four maximally entangled groups of $|\Phi^\pm\rangle$, $|\Psi^\pm_1\rangle$, $|\Psi^\pm_2\rangle$, and $|\Psi^\pm_3\rangle$.

\begin{figure}
	\centering
	\includegraphics[width=0.9\linewidth]{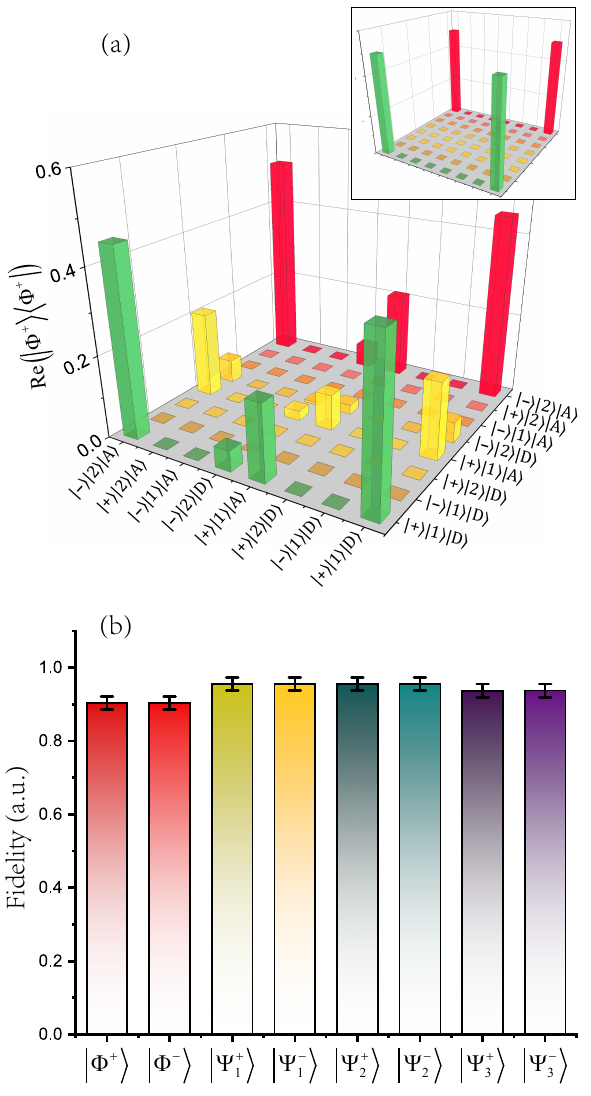}
	\caption{\footnotesize \textbf{GHZ state tomography.} (a) The theoretical (inset) and experimental density matrix of the first GHZ state, $|\Phi^+\rangle$. From the density matrix of each state, the fidelity was calculated, shown in (b).  Error bars show the standard error.}
	\label{f5}
\end{figure}

To quantitatively infer the fidelity of our classically entangled GHZ states, we introduce a new tomography based on Bell state projections, with the results shown in Fig.~\ref{Bell}. The two lobes of the GHZ state beam are overlapped at the non-OAM planar state, projected onto polarization states, and the visibility in fringes measured: a Bell state projection (see Supplementary Material, section~\ref{AH}, for the theory of this). This allows the state amplitudes and relative phases to be determined, and hence the GHZ states can be clearly distinguished.  From the visibility of each projection in the tomographic measurement, a density matrix can be inferred and the fidelity of each GHZ state quantitatively calculated.  The results of this are shown in Fig.~\ref{f5} (a) for the $|\Phi^+\rangle$ state (as the lowest fidelity example) with the measured fidelities of all the GHZ states shown in Figure~\ref{f5} (b). All other density matrix results can be found in the Supplementary Material, Fig.~\ref{tomo}. We find that the theoretical (inset) and measured density matrices are in very good agreement, with the laser generating GHZ states with fidelities of approximately $90\%$.

\begin{figure*}
	\centering
	\includegraphics[width=\linewidth]{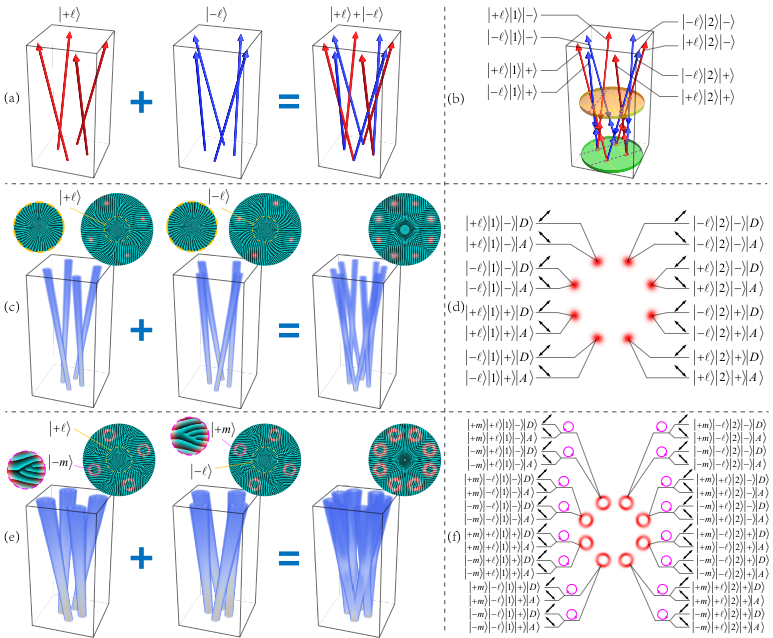}
	\caption{\footnotesize \textbf{Towards higher-dimensional classical entanglement.} Spatial SU(2) geometric beams can carry OAM and the OAM can be a DoF manifesting additional dimensions in classical entanglement. (a) A superposed trajectory on SU(2) state $|\Omega=1/4\rangle$ including the positive-OAM and negative-OAM decomposed SU(2) trajectories, (b) which can exactly fulfill a complete SU(2) oscillation in a degenerate cavity. (c) Based on the ray-wave duality, the corresponding geometric mode can be represented by the superposition of a positive-OAM and a negative-OAM vortex SU(2) beams (the topological phase manifesting the OAM and intensity wave-packet are shown in the inserts). This generalized beam can be expressed into 16-dimensional state with 4 DoFs and 16 eigenstates noted in figure (d). (e) An exotic multi-vortex SU(2)-structured mode is obtained by the superposition of two multi-vortex SU(2) modes with opposite main-OAMs, and the sub-OAM carried by each sub-ray mode can also play as a new DoF. This generalized beam can be expressed into 32-dimensional state with 5 DoFs and 32 eigenstates noted in figure (f).}
	\label{HO}
\end{figure*}

\vspace{0.5cm}
\noindent \textbf{Towards higher dimensions.}  So far we have demonstrated 8-dimensional 3-partite GHZ states by SU(2) structured light. Nevertheless, the SU(2) structured light has the potential to realize higher-dimensional multi-partite states. For instance, we can use a superposed OAM state, as shown in the Fig.~\ref{HO} (a), where the superposed SU(2) trajectory includes the positive and negative twisted decomposed SU(2) trajectories. More importantly, the superposed trajectory fulfills the condition of a complete oscillation in a degenerate cavity, as illustrated in Fig.~\ref{HO} (b), so that it can be directly generated from the laser (see Supplementary Material). Therefore, the corresponding ray-wave beam introducing OAM as a new DoF shown in Fig.~\ref{HO} (c), can be expressed as a 16-dimensional and 4-partite classically entangled state, with the eigenstates shown in Fig.~\ref{HO} (d). For exploring even more DoFs, we can replace the Gaussian beams along the SU(2) orbits in a geometric mode based on ray-wave duality. In this generalized beam, there is not only a main OAM state $|\pm\ell\rangle$ along the propagation axis, but also OAM $|\pm\rangle$ carried by the sub-vortex beams along the SU(2) ray-orbits, as shown in Fig.~\ref{HO} (e). This exotic SU(2) beam also satisfies the conditions of an SU(2) coherent state, realizing a 32-dimensional and 5-partite state, as noted in Fig.~\ref{HO} (f). See Supplementary Material, section~\ref{AK}, for more theoretical descriptions on these states.

\section*{Discussion and conclusion}
Here we have demonstrated a simple laser that can produce exotic vectorial states of light that are non-separable in high-dimensions, beyond the existing laser limitation of two dimensions in polarization structured vector beams.  Our cavity has no internal elements and achieves mode control by simple adjustments to the cavity (e.g., length and pump position/power).  We have introduced a new tomography approach for verifying high-dimensional states, which we used to quantify the fidelity of our GHZ states.  In our experimental demonstration we have used a cavity operating at the degenerate state of $|\Omega=1/4\rangle$ for 8 dimensions and simulating three partite entanglement.  By operating at a more general degenerate state, the general SU(2) modes would be extended to a higher-dimensional space, which could be extended even further by replacing the Gaussian-like orbital modes by more exotic structures~\cite{lu2011generation,tuan2018characterization}, including those based on OAM (as described in the previous section).  While we have tailored our laser modes to create the first classically entangled GHZ states, other states could be tailored such as $W$ \cite{dur2000three}, NOON \cite{kok2002creation} and CAT \cite{liu2019classical} states, or our high-dimensional laser could be used as the input to alternative path modulation schemes \cite{fickler2014interface,pabon2019high}.  

Our work allows for a compact and simple generation scheme to produce new states of structured light directly from the source.  In particular, we demonstrate classical analogues to multi-partite high-dimensional quantum states, which have already been suggested for tasks such as quantum channel error estimation and correction, super-resolution imaging, metrology and sensing, optical communication, and quantum decoherence studies under easily controlled conditions \cite{aiello2015quantum,mabena2017high,mabena2020quantum,korolkova2019quantum,forbes2019quantum}. Finally, in addition to the practical laser device we also offer a complete theoretical framework for our vectorially structured light beams, providing a rich parameter space for extending fundamental studies with structured light.

\section*{Methods}

\noindent \textbf{Creating GHZ states.}    The GHZ generation step is shown as part of Fig.~\ref{f3}.  The relevant transformations for generating our GHZ states are executed in two consecutive steps. First through intensity modulation then followed by dynamic phase (SLM) modulation. A graphical representation of the scheme is illustrated in Fig.~\ref{f3}~(a), showing the intensity modulation of the orbits (rows) and polarization control (columns). Each panel describes the required modulation to produce each of the desired GHZ states.

To achieve the path modulation, we use an iris strategically positioned at I$_1$ or I$_2$. We unpack this technique in Fig.~\ref{f3}~(b) showing an example for modulation at location I$_1$. The four quadrants represent the full orbits while the grey regions represent the filtered orbits. Due to the special spatial structure of the SU(2) geometric beam, the iris induces a path-dependent intensity modulation: when an iris with an appropriate aperture size is placed at the negative Rayleigh length position (I$_1$ position), paths $\ket{-}\ket{1}$ and $\ket{+}\ket{2}$ can be blocked, resulting in a diagonal intensity pattern in the corresponding vortex SU(2) beam [see Fig. \ref{f3} (b)]. When applied at the positive Rayleigh length position (I$_2$ position), $\ket{+}\ket{1}$ and $\ket{-}\ket{2}$ are blocked, resulting in an anti-diagonal intensity pattern of vortex SU(2) beam. 

For polarization modulation, we first convert all rays to circular polarization with a QWP and direct them towards the SLM with controlled location. This is illustrated graphically in Fig.~\ref{f3}~(c), where the four orbits are marked with (anti-)diagonal polarization states shows as arrows. The phase mask of the SLM is split into two parts, one to modulate the states $\ket{1}$ and the other to modulate states $\ket{2}$.  Since the beam is circularly polarized and the SLM is only sensitive to horizontal component, the polarization of the $\ket{1}$ and $\ket{2}$ states can be independently controlled by encoding phase retardations on each section of the SLM. When the encoded phase-step mask is set to $\pi/2$ and $3\pi/2$ for the split screen, respectively, the states $\ket{1}\ket{D}$ and $\ket{2}\ket{A}$ are produced.  Likewise, when the mask is flipped ($3\pi/2$ and $\pi/2$) the states $\ket{1}\ket{A}$ and $\ket{2}\ket{D}$ are produced. The combination of intensity and polarization modulation results into the four kinds of vector vortex beams corresponding to the four maximally entangled groups of three-partite GHZ states. Finally, the inter-modal phase between states is easily controlled with a thin BK7 plate that is partially inserted into the path of one ray group to add a phase difference 0 or $\pi$.
\vspace{0.5cm}

\noindent \textbf{Bell state projections.} We describe the working principle of our Bell state measurement approach, shown as part of Fig.~\ref{f3}, comprised of a $F=250$~mm focusing lens, polarizer and CCD camera. The camera can be placed at different regions within the focus of the lens.

Our measurement scheme exploits the properties of the multi-particle GHZ states. A GHZ state will be reduced to a two particle Bell state after a superposition measurement of one of the particles (See Supplementary Materials, section~\ref{AG}). For classical GHZ states, superposition projections of one of the DoFs leaves the remaining DoFs to collapse to maximally entangled Bell states (See Supplementary Material, section~\ref{AH}, for the full explanation). Here the polarization DoF of freedom is chosen as a candidate to realize the Bell projection.

After projections onto $|H\rangle$ and $|V\rangle$ states, $|\Phi^\pm\rangle$ or $|\Psi^\pm_3\rangle$ states would be reduced to $|\psi^\pm\rangle$ and $|\psi^\mp\rangle$ states, $|\Psi^\pm_1\rangle$ or $|\Psi^\pm_2\rangle$ to $|\phi^\pm\rangle$ and $|\phi^\mp\rangle$ [refer to Eqs. (\ref{p})-(\ref{cb2}), Supplementary Material]. The ``$+$'' and ``$-$'' signals in the Bell states can be distinguished by the complementary interferometric fringes of the corresponding phase difference of 0 and $\pi$ between two SU(2) ray-orbits. For measuring $|\Phi^\pm\rangle$ and $|\Psi^\pm_3\rangle$, the CCD camera can be located at $z=-z_R$ position where $|-1\rangle$ and $|+2\rangle$ orbits are overlapped. For $|\Psi^\pm_1\rangle$ and $|\Psi^\pm_2\rangle$, the CCD camera should be located at $z=z_R$ position where $|+1\rangle$ and $|-2\rangle$ orbits are overlapped. Without the polarization projection, the pattern shows no fringes since the light on the corresponding two orbits is incoherent. After projection on $|H\rangle$ or $|V\rangle$ states, different interference fringes would be observed for different reduced Bell states. 

For the group $|\Phi^\pm\rangle$, the ``$\pm$'' cannot be distinguished by the intensity patterns. If we project the polarization onto $|H\rangle$ state to observe the pattern of $\langle H|\Phi^\pm\rangle$, pattern of original state $|\Phi^\pm\rangle$ will be reduced into Bell states $|\psi^\pm\rangle$ and two different patterns of complementary fringes will be observed, center-bright fringes for $|\psi^+\rangle$ (the original state should be $|\Phi^+\rangle$) and center-dark fringes for $|\psi^-\rangle$ (the original state should be $|\Phi^-\rangle$). We can also use projected state $\langle V|\Phi^\pm\rangle$ to distinguish the ``$\pm$'' that $\langle V|\Phi^+\rangle$ should be center-dark fringes corresponding to Bell state $|\psi^-\rangle$ and $\langle V|\Phi^-\rangle$ should be center-bright fringes corresponding to Bell state $|\psi^+\rangle$. In the experiment, we can use a BK7 thin plate to cover one of the two orbits and rotate slightly to control the phase difference between them to control a phase difference of $\pi$, switching from ``$+$'' to ``$-$'' state. Other GHZ states can be generated by the similar way fulfilling a completed set in 8-dimensional Hilbert space.

\section*{Acknowledgements}
National Key Research and Development Program of China (2017YFB1104500); National Natural Science Foundation of China (61975087); Natural Science Foundation of Beijing Municipality (4172030); Beijing Young Talents Support Project (2017000020124G044). 

\section*{Authors' contribution}
Y.S. and A.F. proposed the idea and conceived the experiment, Y.S. performed experiments and the theory, Y.S. and I.N. proposed the projection method of GHZ states, Y.S and X.Y. performed the numerical simulations, D.N. helped with the SLM modulation experiments, all authors contributed to data analysis and writing the manuscript, and the project was supervised by A.F. and M.G.

\section*{Competing financial interests}
The authors declare no financial or competing interests.

\section*{Materials and correspondence}
Correspondence and requests for materials should be addressed to Y.S. (shenyj15@tsinghua.org.cn) and A.F. (Andrew.Forbes@wits.ac.za).

%\begin{acknowledgments}
%Dr.~Yijie Shen would like to thank Prof.~Xing Fu in Tsinghua university for useful discussions.
%\end{acknowledgments}

\clearpage
\newpage

%\nocite{*}

\providecommand{\noopsort}[1]{}\providecommand{\singleletter}[1]{#1}%

\clearpage
\newpage

\section*{Supplementary Information}
\renewcommand{\theequation}{S.\arabic{equation}}
\beginsupplement{
\subsection{Schr{\"o}dinger coherent state}\label{AA}
A coherent state is a specific quantum state whose behavior most closely resembles the classical state, and particularly adapted for studying the quantum-to-classical transition, where the quantum probability wave-packet can be coupled with classical movement~\cite{perelomov2012generalized}. For a simple 1-D linear oscillator, the Hamiltonian is given by:
\begin{align}
\nonumber
\mathcal{H}&=\frac{1}{2m}p^2+\frac{1}{2}m\omega^2x^2\\&={\frac{1}{2}\left( a^{\dagger }{{a}}+{{a}}a^{\dagger } \right)\hbar {{\omega }}}={\left( a^{\dagger }{{a}}+\frac{1}{2} \right)\hbar {{\omega }}},
\end{align}
where $m$ is the mass, $\omega$ is the unperturbed frequency, $p$ is the momentum operator, $x$ is the coordinate operator, ${a}^{\dagger}$ and ${a}$ are the ladder (creation and annihilation) operators of photon, and $\hbar$ is the reduced Planck constant. The eigenstate under coordinate representation is expressed by Hermite function:
\begin{equation}
|n\rangle=\sqrt{\frac{1}{\sqrt{\pi}2^nn!}}H_n(\xi)\text{e}^{-\xi^2/2},
\label{n}
\end{equation}
where $\xi=\sqrt{m\omega/\hbar}\cdot x$, with eigenvalue of:
\begin{equation}
E_n=\left(n+\frac{1}{2}\right)\hbar {{\omega }},\label{E}
\end{equation}
According to the definition of coherent state, the coherent state under coordinate representation, i.e. Schr{\"o}dinger coherent state~\cite{schrodinger1926stetige}, is given by:
\begin{equation}
|\alpha\rangle ={{\text{e}}^{-{{{\alpha }^{*}}\alpha }/{2}\;}}\sum\limits_{n=0}^{\infty }{\frac{{{\alpha }^{n}}}{\sqrt{n!}}}\left|n\right\rangle {{\text{e}}^{-{\text{i}{{E}_{n}}t}/{\hbar }\;}}.\label{S}
\end{equation}
Substituting Eq.~(\ref{n}) and Eq.~(\ref{E}) into Eq.~(\ref{S}) and applying the generating function of Hermite polynomials, $\exp(2xt-t^2)$=$\sum_{n=0}^{\infty}H_n(x)t^n/n!$, we get:
\begin{align}
\nonumber
& \left| \alpha  \right\rangle ={{\text{e}}^{-\frac{{{\left| \alpha  \right|}^{2}}}{\text{2}}}}\sum\limits_{n=0}^{\infty }{\frac{{{\left( \left| \alpha  \right|{{\text{e}}^{\text{i}\delta }} \right)}^{n}}}{\sqrt{n!}}}{\frac{{H}_{n}\left( \xi  \right)}{\sqrt{\sqrt{\pi }{{2}^{n}}n!}}}{{\text{e}}^{-\frac{{{\xi }^{2}}}{2}}}{{\text{e}}^{-\text{i}\left( n+\frac{\text{1}}{\text{2}} \right)\omega t}} \\ \nonumber
& =\frac{1}{\sqrt{\sqrt{\pi }}}{{\text{e}}^{-\frac{{{\left| \alpha  \right|}^{2}}+{{\xi }^{2}}}{\text{2}}}}{{\text{e}}^{-\frac{\text{i}\omega t}{\text{2}}}}\sum\limits_{n=0}^{\infty }{\frac{{{\left[ \frac{\left| \alpha  \right|{{\text{e}}^{-\text{i}\left( \omega t-\delta  \right)}}}{\sqrt{2}}\; \right]}^{n}}{{H}_{n}}\left( \xi  \right)}{n!}} \\ 
& =\frac{1}{\sqrt{\sqrt{\pi }}}{{\text{e}}^{-\frac{{{\left| \alpha  \right|}^{2}}+{{\xi }^{2}}}{\text{2}}}}{{\text{e}}^{-\frac{\text{i}\omega t}{\text{2}}}}{{\operatorname{e}}^{-\frac{{{\left| \alpha  \right|}^{2}}{{\text{e}}^{-\text{i2}\left( \omega t-\delta  \right)}}}{2}+\sqrt{2}\left| \alpha  \right|{{\text{e}}^{-\text{i}\left( \omega t-\delta  \right)}}\xi }},  
\end{align}
where $\delta$ is the argument of $\alpha$. Then, the probability wave-packet of Schr{\"o}dinger coherent state can be derived by:
\begin{align}
\nonumber
\langle \alpha | \alpha \rangle &=\frac{1}{\sqrt{\pi }}{{\text{e}}^{-{{\left| \alpha  \right|}^{2}}-{{\xi }^{2}}}}{{\operatorname{e}}^{-{{\left| \alpha  \right|}^{2}}\text{cos}\left[ \text{2}\left( \omega t-\delta  \right) \right]+2\sqrt{2}\left| \alpha  \right|\xi \cos \left( \omega t-\delta  \right)}} \\ \nonumber
& =\frac{1}{\sqrt{\pi }}{{\operatorname{e}}^{-{{\xi }^{2}}-2{{\left| \alpha  \right|}^{2}}\text{co}{{\text{s}}^{2}}\left( \omega t-\delta  \right)+2\sqrt{2}\left| \alpha  \right|\xi \cos \left( \omega t-\delta \right)}} \\ 
&=\frac{1}{\sqrt{\pi }}{{\operatorname{e}}^{-{{\left[ \xi -\sqrt{2}\left| \alpha  \right|\cos \left( \omega t-\delta \right) \right]}^{2}}}}.
\end{align}
As shown in Fig.~\ref{cs}, the peak of wave-packet of is along the trajectory of corresponding classical oscillator, i.e. $\xi =\sqrt{2}\left| \alpha  \right|\cos \left( \omega t-\delta \right)$, manifesting the quantum-classical coupling.
\begin{figure*}
	\centering
	\includegraphics[width=0.75\linewidth]{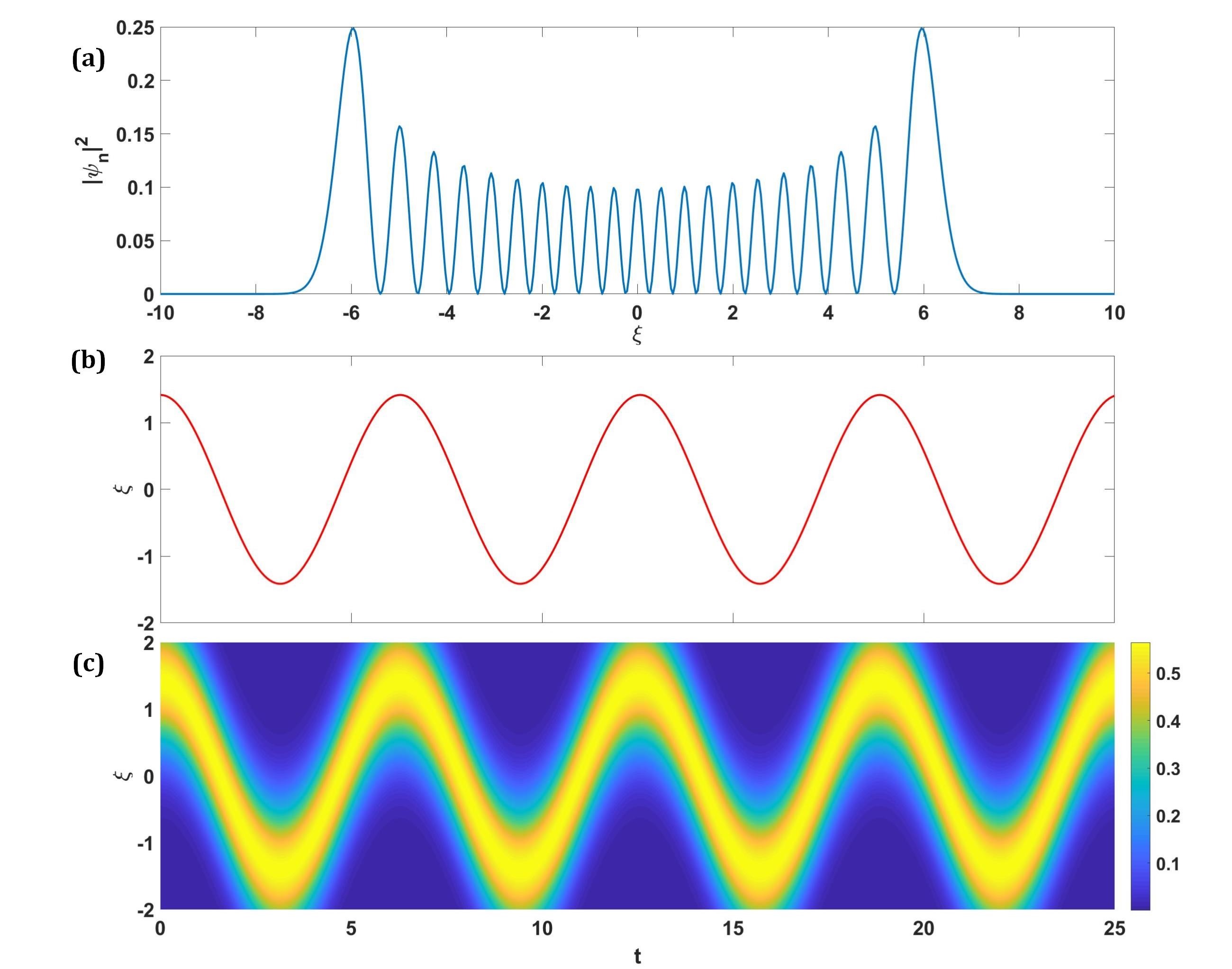}
	\caption{\footnotesize \textbf{Schr{\"o}dinger coherent state.}  (a) Probability wave-packet of eigenstate of 1-D linear oscillator ($n$=20). (b) Trajectory of the classical movement of 1-D linear oscillator and (c) the probability wave-packet of coherent state ($\alpha=1$, $\omega=1$, $\delta=0$).}
	\label{cs}
\end{figure*}

\subsection{SU(2) coherent state}\label{AB}
In quantum optics, the Hamiltonian for the 3-D linear harmonic oscillator is given by:
\begin{equation}
\mathcal{H}=\sum\limits_{j}{\frac{1}{2}\left( a_{j}^{\dagger }{{a}_{j}}+{{a}_{j}}a_{j}^{\dagger } \right)\hbar {{\omega }_{j}}}=\sum\limits_{j}{\left( a_{j}^{\dagger }{{a}_{j}}+\frac{1}{2} \right)\hbar {{\omega }_{j}}},
\end{equation}
where $j=x,y,z$. The generalized Hamiltonian related to SU(2) transformations for the coupled oscillator systems can be modeled as:
\begin{equation}
\mathcal{H}=\mathcal{H}_0+\sum\limits_{j}{\Omega_jJ_j},
\label{Hsu2}
\end{equation}
where $\mathcal{H}_0=(a_xa_x^\dagger+a_ya_y^\dagger+1)\hbar\omega_0$ is the Hamiltonian for the 2-D isotropic oscillator deciding the transverse wave-packet on $(x,y)$ plane.  The coupling parameters $\Omega_j$ are assumed to be real
constants, and the operators under Schwinger representation reveal the SU(2)-Lie group accommodating two linear oscillators and an angular momentum oscillator~\cite{buvzek1989generalized,wodkiewicz1985coherent}. 
\begin{equation}
\left\{
	\begin{aligned}
	{{J}_{x}}&=\frac{1}{2}\left( a_{x}^{\dagger }{{a}_{y}}+a_{y}^{\dagger }{{a}_{x}} \right) \\
	{{J}_{y}}&=\frac{-\text{i}}{2}\left( a_{x}^{\dagger }{{a}_{y}}-a_{y}^{\dagger }{{a}_{x}} \right) \\
	{{J}_{z}}&=\frac{1}{2}\left( a_{x}^{\dagger }{{a}_{x}}-a_{y}^{\dagger }{{a}_{y}} \right)
	\end{aligned}
	\right.
\end{equation} 
Operators $J_j$ satisfy the SU(2)-Lie commutator algebra $[J_i,J_j] = \text{i}\varepsilon_{i,j,k}J_k$ ($\{i,j,k\}=\{1,2,3\}=\{x,y,z\}$),
where the Levi-Civita tensor $\varepsilon_{i,j,k}$ is equal to $+1$ and $-1$ for even and odd permutations of its indices, respectively, and zero otherwise. The Hamiltonian in Eq.~(\ref{Hsu2}) can not only represent a host of entanglement mechanisms~\cite{sanders1995optical,kim2002entanglement} but also be associated with astigmatism and aberration in wave optics, relevant in high-order laser pattern formations~\cite{agarwal19992,su21}. For the statistics of quantum number at transverse oscillation, SU(2) coherent state is defined as:
\begin{equation}
\left| \alpha  \right\rangle =\exp \left( \alpha {{J}_{+}}-{{\alpha }^{*}}{{J}_{-}} \right)\left|j,-j \right\rangle,
\label{alf}
\end{equation}
where ${{J}_{\pm }}={{J}_{x}}\pm \text{i}{{J}_{y}}$ are the ladder (creation and annihilation) operators of angular momentum, and $j$ is a certain integer or half-integer. Using the disentangling theorem for angular-momentum operators~\cite{wodkiewicz1985coherent}, we can rewrite Eq.~(\ref{alf}) in the following form:
\begin{equation}
\left| \tau  \right\rangle ={{\left( 1+{{\left| \tau  \right|}^{2}} \right)}^{-j}}\exp \left( \tau {{J}_{+}} \right)\left|j,-j\right\rangle,
\label{tau}
\end{equation}
Hereinafter, we express Eq.~(\ref{tau}) into eigenstates representation via unitary transformation. According to Taylor expansion, the exponential operator in  Eq.~(\ref{tau}) can be expanded as:
\begin{equation}
\exp \left( \tau {{J}_{+}} \right)=\sum\limits_{n=0}^{\infty }{\frac{{{\left( \tau {{J}_{+}} \right)}^{n}}}{n!}}.
\label{Texp}
\end{equation}
Substitute Eq.~(\ref{Texp}) into Eq.~(\ref{tau}) and apply unitary transformation into angular-momentum representation:
\begin{align}
\nonumber
 \left| \tau  \right\rangle &={{\left( 1+{{\left| \tau  \right|}^{2}} \right)}^{-j}}\sum\limits_{k=-j}^{j}{\left\langle j,k \right|\exp \left( \tau {{J}_{+}} \right)\left|j,-j \right\rangle \left| j,k \right\rangle } \\ \nonumber 
& ={{\left( 1+{{\left| \tau  \right|}^{2}} \right)}^{-j}}\sum\limits_{k=-j}^{j}{\left\langle j,k \right|\sum\limits_{n=0}^{\infty }{\frac{{{\left( \tau {{J}_{+}} \right)}^{n}}}{n!}}\left|j,-j\right\rangle \left| j,k \right\rangle } \\ 
& ={{\left( 1+{{\left| \tau  \right|}^{2}} \right)}^{-j}}\sum\limits_{k=-j}^{j}{\left\langle j,k \right|\frac{{{\left( \tau {{J}_{+}} \right)}^{j+k}}}{\left( j+k \right)!}\left| j,-j \right\rangle\left| j,k \right\rangle }.
\label{tau2}
\end{align}
According to the property of ladder operators:
\begin{equation}
{{J}_{\pm }}\left| j,k \right\rangle =\sqrt{j\left( j+1 \right)-k\left(k\pm 1 \right)}\hbar \left| j,k\pm 1 \right\rangle,
\end{equation}
Eq.~\ref{tau2} can be rewritten as:
\begin{align}
\nonumber
\left| \tau  \right\rangle &=\frac{\sum\limits_{k=-j}^{j}{\frac{{\underset{i=-j}{\overset{s-1}{\mathop \prod }}\sqrt{ j\left( j+1 \right)-i(i+1) }}}{\left( j+k \right)!}{{\tau }^{j+k}}\left| j,k \right\rangle }}{{\left( 1+{{\left| \tau  \right|}^{2}} \right)}^{j}} \\ 
& ={{\left( 1+{{\left| \tau  \right|}^{2}} \right)}^{-j}}\sum\limits_{k=-j}^{j}{{{\left( \begin{matrix}
			2j  \\
			j+k  \\
			\end{matrix} \right)}^{{1}/{2}\;}}{{\tau }^{j+k}}\left| j,k \right\rangle }, 
\end{align}
After substituting $N=2j$ and $K=j+k$ ($N$ is a constant integer, $K$ is integer yielded $0\leqslant K\leqslant N$),  we get:
\begin{equation}
\left| \tau  \right\rangle ={{\left( 1+{{\left| \tau  \right|}^{2}} \right)}^{-{N}/{2}\;}}\sum\limits_{K=0}^{N}{{{\left( \begin{matrix}
				N  \\
				K  \\
			\end{matrix} \right)}^{{1}/{2}}}{{\tau }^{K}}\left| K,N \right\rangle }.
		\label{tau3}
\end{equation}
In another usually used form, $\tau$ is rewritten as the normalized argument form $\tau={{\text{e}}^{\text{i}\phi}}$, and Eq.~(\ref{tau3}) is rewritten as the phase state:
\begin{equation}
\left| \phi  \right\rangle =\frac{1}{{{2}^{{N}/{2}\;}}}\sum\limits_{K=0}^{N}{{{\left( \begin{matrix}
				N  \\
				K  \\
			\end{matrix} \right)}^{{1}/{2}\;}}{{\text{e}}^{\text{i}K\phi }}\left| K,N \right\rangle },
		\label{phi}
\end{equation}
where the eigenstates $| K,N \rangle$ should fulfill the orthogonality $\left\langle K,N| L,N \right\rangle ={\delta_{KL}}$, where $\delta_{i,j}$ is the Kronecker delta, and the completeness $a_{x}^{\dagger }{{a}_{x}}\left| K,N \right\rangle =K\left| K,N \right\rangle$, $a_{y}^{\dagger }{{a}_{y}}\left| K,N \right\rangle =\left( N-K \right)\left| K,N \right\rangle$, $\sum\limits_{K=0}^{N}{\left| K,N \right\rangle \left\langle  K,N \right|}=1$.

\begin{figure*}
	\centering
	\includegraphics[width=\linewidth]{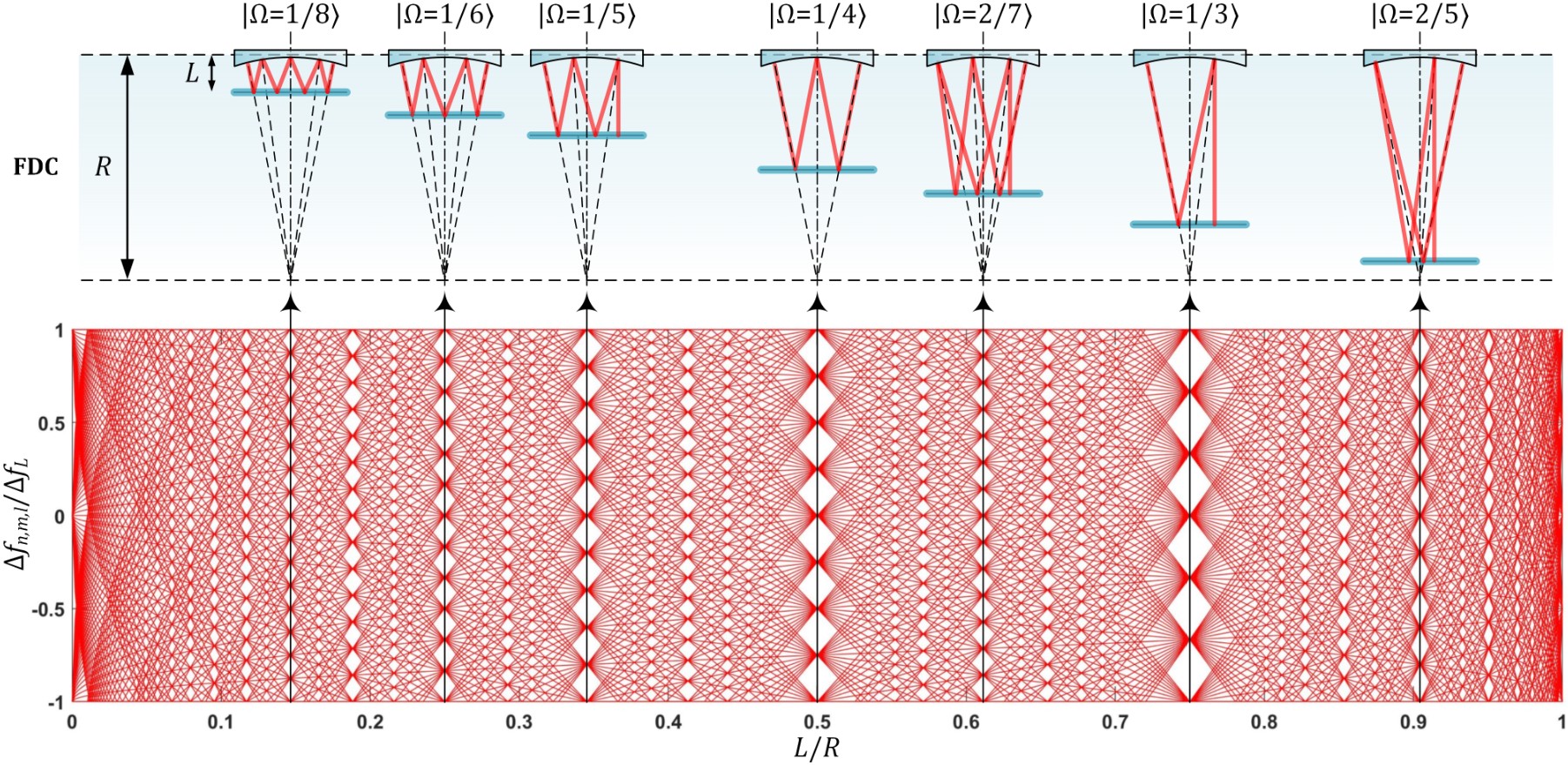}
	\caption{\footnotesize \textbf{Frequency-degenerate spectrum and the ray representation in laser cavity.}  The frequency-degenerate spectrum $(f_{n,m,l}-f_{n_0,m_0,l_0})/\Delta f_L$ of the ideal spherical cavity as a function of the normalized cavity length $L/R$ for the range of $|n-n_0|\le12$, $|m-m_0|\le12$, and $|l-l_0|\le12$, where some degeneracy states $|\Omega=P/Q\rangle$ are marked at corresponding positions with corresponding schematics of ray representation of SU(2) oscillation.}
	\label{spect}
\end{figure*}

\subsection{Frequency-degenerate state}\label{AC}
For realizing SU(2) coherent state in a laser cavity, the eigenstates should be the eigenmodes of the resonator and fulfill the coherent-superposition condition of SU(2) wave-packet. Without loss of generality, we consider a plano-concave cavity with the length of $L$, formed by a gain medium, a concave spherical mirror with the radius of curvature of $R$ as the output coupler, and a plane mirror high-reflective for laser. The eigenmodes ${\psi }_{n,m,l}$ ($n,m$ are the indices of transverse mode and $l$ is the index of longitudinal mode) and the eigenvalues ${k}_{n,m,l}$ for a laser cavity can be solved from the Helmholtz equation:
\begin{equation}
\left( {{\nabla }^{2}}\text{+}k_{n,m,l}^{2} \right){{\psi }_{n,m,l}}\left( x,y,z \right)=0.
\end{equation}
Under the paraxial approximation, the eigenmodes being separable in Cartesian coordinate can be expressed as Hermite-Gaussian (HG) modes:
\begin{align}
\nonumber
\psi _{n,m,l}^{\left( \text{HG} \right)}\left( x,y,z \right)=&\frac{1}{\sqrt{{{2}^{m+n-1}}\pi m!n!}}\frac{1}{w\left( z \right)}\text{e}^{ -\frac{{{x}^{2}}+{{y}^{2}}}{{{w}^{2}}\left( z \right)}}\\\nonumber
&\times H_n\left[\frac{\sqrt{2}x}{w\left( z \right)} \right]H_m\left[ \frac{\sqrt{2}y}{w\left( z \right)} \right]\\
&\times\text{e}^{\text{i}{{k}_{n,m,l}}\widetilde{z}-\text{i}\left( m+n+1 \right){{\vartheta }}\left( z \right) },
\label{hg}
\end{align}
where $\vartheta(z)=\tan^{-1}(z/z_R)$ is the Gouy phase, $H_n(\cdot)$ represents the Hermite polynomials of $n$-th order, $k_{n,m,l}=2\pi f_{n,m,l}/c$, $f_{n,m,l}$ is the eigenmode frequency, $c$ is the speed of light, $\widetilde{z}=z+(x^2+y^2)z/[2(z^2+z_R^2)]$, $w(z)=w_0\sqrt{1+(z/z_R )^2}$, $w_0=\sqrt{(\lambda z_R)/\pi}$ is the beam radius parameter, and $\lambda$ is the emission wavelength. The eigenmode frequency of resonator is given by~\cite{chen2013exploring,tung2016fractal}:
\begin{align}
\nonumber
f_{n,m,l}&=l\Delta f_L+(n+\frac{1}{2})\Delta f_x+(m+\frac{1}{2})\Delta f_y\\
&=\left[l+(n+m+1)\Omega\right]\Delta f_L,
\label{f}
\end{align}
where the longitudinal mode spacing $\Delta f_L=c/(2L)$, here the minor disparity between the physical length and the geometric length is neglected. Without consideration of symmetry breaking, the transverse mode spacing should be $\Delta f_x+\Delta f_y=\Delta f_T=\Delta f_L\vartheta(L)/\pi$. The mode-spacing ratio $\Omega=P/Q=(1/\pi)\cos^{-1}(\sqrt{1-L/R})$ reveals the degeneracy, which is varied in the range between 0 and 1/2 by changing the cavity length as $0<L<R$. The frequency difference in the neighborhood of the indices $(n_0,m_0,l_0)$ is given by $(f_{n,m,l}-f_{n_0,m_0,l_0})/\Delta f_L$, which can illustrate the various degeneracy states distribution as topological joints in the fractal spectrum~\cite{tung2016fractal}. Fig.~\ref{spect} depicts a diagram of the frequency-degenerate spectrum where some degeneracy states $|\Omega=P/Q\rangle$ are marked at corresponding positions. In order to fulfill the condition of coherent superposition, the frequency of every decomposed eigenmodes should be a constant, which requires a coupling effect between transverse and longitudinal modes. If the transverse mode at $x$-axis and longitudinal mode are coupled, i.e. $n+m+l=N$, we can choose a frequency-degenerate family of HG modes as the complete set of orthogonal bases:
\begin{equation}
\left| K,N \right\rangle =\psi _{{{n}_{0}}+QK,m_0,{l_{0}}-PK}^{\left( \text{HG} \right)}\left( x,y,z \right),
\end{equation}
where $n_0$, $m_0$, and $l_0$ are constants with $n_0+m_0+l_0=N$, thus the constant frequency should be:
\begin{align}
\nonumber
f_{n,m,l}&=\left[{{l}_{0}}-PK+({{n}_{0}}+QK+m_0+1)\Omega\right]\Delta f_L\\ \nonumber
&=\left[l_0+(n_0+m_0+1)\Omega\right]\Delta f_L\\
&=f_{n_0,m_0,l_0},
\end{align}
and the corresponding laser wave-packet of SU(2) coherent state is given by:
\begin{align}
\nonumber
\Psi _{{n_0,m_0}}^{N,\Omega,\phi}\left( x,y,z\right)=&\frac{1}{{{2}^{N/{2}}}}\sum\limits_{K=0}^{N}{{\left( \begin{matrix}
		N  \\
		K  \\
		\end{matrix} \right)}^{{1}/{2}}} {{\text{e}}^{\text{i}K{\phi}}}\\
	 &\times\psi _{{{n}_{0}}+QK,m_0,{{l}_{0}}-PK}^{\left( \text{HG} \right)}\left( x,y,z \right),
\label{pgm}
\end{align}
sharing the same form of Eq.~(\ref{phi}). Here we already proved that frequency-degenerate state of laser cavity fulfills the condition for generating a laser wave-packet as SU(2) coherent state. We can also use Laguerre-Gaussian (LG) modes being separable in circular coordinate as the eigenmodes to generate SU(2) vortex beams:
\begin{align}
\nonumber
\Phi _{{n_0,m_0}}^{N,\Omega,\phi}\left( x,y,z\right)=&\frac{1}{{{2}^{N/{2}}}}\sum\limits_{K=0}^{N}{{\left( \begin{matrix}
		N  \\
		K  \\
		\end{matrix} \right)}^{{1}/{2}}} {{\text{e}}^{\text{i}K{\phi}}}\\
&\times\varphi _{{{n}_{0}}+QK,m_0,{{l}_{0}}-PK}^{\left( \text{LG} \right)}\left( x,y,z \right),
\label{vgm}
\end{align}
where the LG modes are given by:
\begin{align}
\nonumber
\varphi _{n,m,l}^{\left( \text{LG} \right)}\left( \rho,\theta,z \right)=&\sqrt{\frac{2p!}{\pi \left( p+\left| \ell \right| \right)!}} \frac{1}{w\left( z \right)}{{\left[ \frac{\sqrt{2}r}{w\left( z \right)} \right]}^{\left| \ell  \right|}}\text{e}^ {-\frac{{{r}^{2}}}{{{w}^{2}}\left( z \right)}}\text{e}^{\text{i}\ell \theta}\\
&\times L_{p}^{\left| \ell  \right|}\left[ \frac{2{{r}^{2}}}{{{w}^{2}}\left( z \right)} \right]\text{e}^{\text{i}{{k}_{n,m,l}}\widetilde{z}-\text{i}\left( m+n+1 \right){{\vartheta }}\left( z \right) },
\end{align}
where $p=\min \left( m,n \right)$, $\ell =\pm \left( m-n \right)$, and $L_{p}^{\ell}\left( \cdot  \right)$ represents the associated Laguerre polynomial with radial and azimuthal indices of $p$ and $\ell$. For $m_0=0$, Fig.~6(c) in the main text shows the 3-D simulation of SU(2) vortex beams with positive, negative, and superposed OAMs, and the inserts show the topological phases of the SU(2) vortex beams. For $m_0\ge1$, the SU(2) beams manifest the multi-LG vortex beams~\cite{lu2011generation,tuan2018characterization}, as shown in Fig.~6(e) in the main text for 3-D simulation together with topological phase, where the main OAM at the center is decided by the index $n_0$ and the sub-OAM carried by sub-LG beams is decided by the index $m_0$. For constituting a completed oscillation in cavity, the positive and negative oscillations should be superposed together forming a  standing wave mode, the phase state expression of which is $|+\rangle+|-\rangle=|\phi\rangle+|2\pi-\phi\rangle$~\cite{chen2004wave}. Figure~\ref{orb} shows the ray presentation (a) intensity wave-packet (b-d) of intracavity geometric modes versus the coherent state phase. From larger $n_0$ to smaller one, the wave-packets perform from ray-like cases to wave-like cases. Figure~\ref{nmM} shows the transverse patterns of planar and vortex geometric modes and the topological phase of vortex geometric modes at SU(2) coherent state $|\Omega=1/4\rangle|\phi=\pi\rangle$ with parameters as $z=z_R$, $M=20$, and various $n_0$ and $m_0$, where some cases perform multi-spot shape while some wave fringes unravel the interference among lights on the sub-orbits, which is manifested by the property of ray-wave duality.
\begin{figure*}
	\centering
	\includegraphics[width=0.8\linewidth]{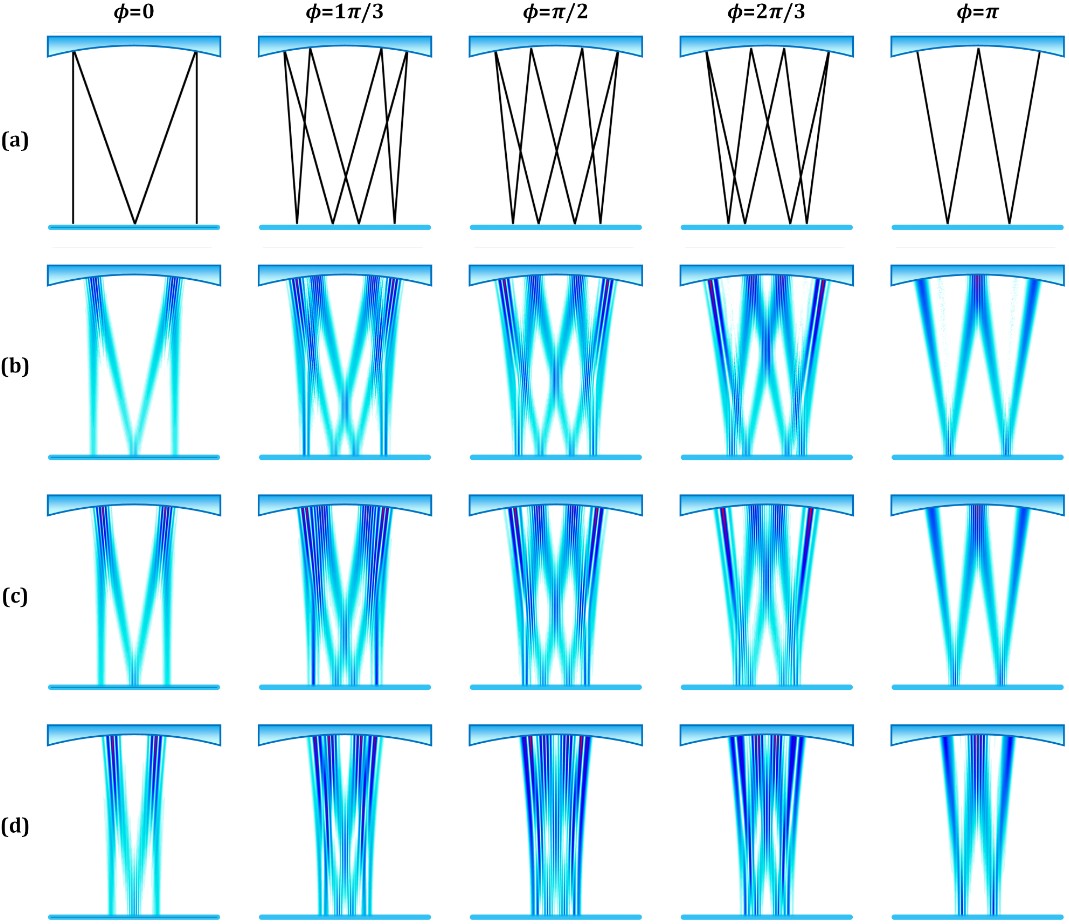}
	\caption{\footnotesize \textbf{Phase states of SU(2) oscillation with ray-wave duality.} Intracavity planar SU(2) geometric mode $|\phi\rangle+|2\pi-\phi\rangle$ oscillating at $(x,z)$-plane at degenerate state $|\Omega=1/4\rangle$: (a) The ray representations and (b-d) intensity wave-packets with $n_0$ from larger to smaller ($n_0=30,20,10$) for various $\phi$. The patterns of wave representations from (b) to (d) change from the case of more ray-like properties to that of more wave-like properties.}
	\label{orb}
\end{figure*}

There are also other ways to realize frequency degeneracy in order to fulfill the coherent superposition of SU(2) wave-packet. For instance:
\begin{equation}
\left| K,N \right\rangle =\psi _{{{n}_{0}}+pK,m_0+qK,{l_{0}}-PK}^{\left( \text{HG} \right)}\left( x,y,z \right),
\label{pq}
\end{equation}
where the integers $p$ and $q$ yield $p+q=Q$, thus the eigenmodes also constitute a frequency-degenerate family with frequency $f_{n_0,m_0,l_0}$. Using the more general Eq.~(\ref{pq}) as the bases of SU(2) coherent state, we can obtain more exotic structured light beams~\cite{su21,tung2016fractal}. 

\begin{figure*}
	\centering
	\includegraphics[width=\linewidth]{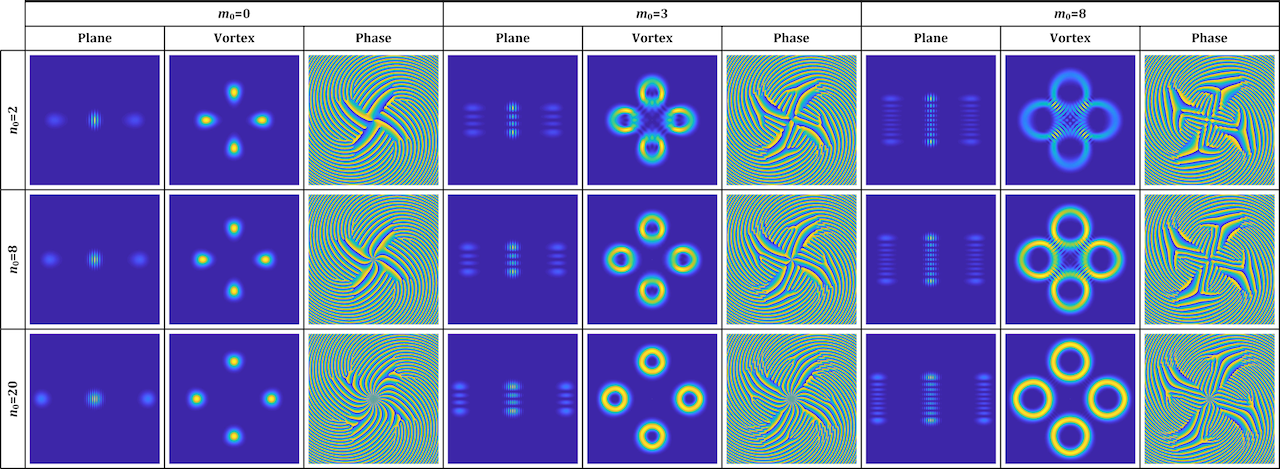}
	\caption{\footnotesize \textbf{Planar and vortex SU(2) geometric modes.} The theoretical intensity transverse patterns of planar and vortex geometric modes and the topological phase of vortex geometric modes at SU(2) coherent state $|\Omega=1/4\rangle|\phi=\pi\rangle$ with parameters as $z=z_R$, $M=20$, and various $n_0$ and $m_0$.}
	\label{nmM}
\end{figure*}

\subsection{Ray-wave duality}\label{AD}
When an optical resonator is operating close to frequency-degenerate state, named frequency-degenerate cavity, its laser mode and intensity would undergo dramatic changes with the principle that laser modes have a preference to be localized on the periodic ray trajectories under off-axis pumping, which is called the ray-wave duality. Like the Schr{\"o}dinger coherent state coupled with the trajectory of classical oscillator, the SU(2) coherent state can also be coupled with the periodic oscillating trajectories in frequency-degenerate cavity.

Based on geometrical optics, the ABCD matrix is used to characterize the propagation property of the optical ray trajectories inside a stable plano-concave cavity. Since the cavity length satisfies $L/R=\sin^2(\Omega\pi)$ under degeneracy state $|\Omega=P/Q\rangle$, the corresponding ABCD matrix of the frequency-degenerate cavity is given by:
\begin{equation}
\mathbf{A}=\left[\begin{matrix}
1-\frac{2L}{R} & 2L\left( 1-\frac{L}{R} \right)  \\
-\frac{2}{R} & 1-\frac{2L}{R}  \\
\end{matrix} \right]=\left[\begin{matrix}
\cos \left( 2\Omega \pi  \right) & \frac{R}{2}{{\sin }^{2}}\left( 2\Omega \pi  \right)  \\
-\frac{2}{R} & \cos \left( 2\Omega \pi  \right)  \\
\end{matrix} \right].
\end{equation}
After $n$ times of round trips in the frequency-degenerate cavity, the matrix is derived as:
\begin{equation}
{{\mathbf{A}}^{n}}=\left[ \begin{matrix}
\cos \left( 2n\Omega \pi  \right) & \frac{R}{2}{{\sin }^{2}}\left( 2n\Omega \pi  \right)  \\
-\frac{2}{R}\frac{\sin \left( 2n\Omega \pi  \right)}{\sin \left( 2\Omega \pi  \right)} & \cos \left( 2n\Omega \pi  \right)  \\
\end{matrix} \right].
\end{equation}
Because $Q\Omega=P$ is an integer, $\cos \left( 2Q\Omega \pi  \right)=1$, $\sin \left( 2Q\Omega \pi  \right)=0$, and $Q$-th power of $\mathbf{A}$ is an unit matrix:
\begin{equation}
{{\mathbf{A}}^{Q}}=\left[\begin{matrix}
\cos \left( 2Q\Omega \pi  \right) & \frac{R}{2}{{\sin }^{2}}\left( 2Q\Omega \pi  \right)  \\
-\frac{2}{R}\frac{\sin \left( 2Q\Omega \pi  \right)}{\sin \left( 2\Omega \pi  \right)} & \cos \left( 2Q\Omega \pi  \right)  \\
\end{matrix} \right]=\mathbf{I}.
\label{a}
\end{equation}
Equation (\ref{a}) reveals that an optical ray oscillating at an arbitrary position within the cavity would coincide exactly with the initial state after $Q$ times of round trips. Therefore, it is proved that the lasing modes have a preference to be localized on the periodic ray trajectories in a frequency-degenerate cavity. The schematics of classical oscillating trajectories at various states $|\Omega=P/Q\rangle$ are shown in Fig.~\ref{spect}. Manifested by the ray matrix, the parametric equation for each periodic orbit in SU(2) oscillation can be derived. For the planar geometric modes, the orbits can be derived as:
\begin{align}
\nonumber
x_{s}^{\pm }(z)&=\sqrt{N}{{w}_{0}}\left[ \cos \left( {{\theta }_{s}}+{{\phi}_x} \right)\mp \left( z/{{z}_{R}} \right)\sin \left( {{\theta }_{s}}+{{\phi}_x} \right) \right] \\ 
& =\sqrt{N}w(z)\cos \left[ {{\theta }_{s}}+{{\phi }_x}\pm \vartheta (z) \right]  
\end{align}
where ${{\theta}_{s}}=(P/Q)2\pi s,s=0,1,2,\cdots, Q-1$ is the running index for the different rays, ${\phi}_x$ is the phase factor related to the initial position and direction, and $+$ and $-$ in the symbol of $\pm$ indicate the backward and forward rays, respectively. Defining the dimensionless variable $\widetilde{x}=\sqrt{2}x/w(z)$, the expression for the ray equation can be expressed as $\widetilde{x}(z)=\operatorname{Re}[\sqrt{2}{u}_{s}^{\pm }(z)]$ with:
\begin{equation}
u_{s}^{\pm }(z)=\sqrt{{N}}\text{e}^{-\text{i}[{{\theta }_{s}}+{{\phi }_x}\pm {{\vartheta }}(z)]}.
\end{equation}
when $\phi_x=n\pi/Q$ ($n\in\mathbb{Z}$), the forward and backward rays would be coincidently overlapped, and the bouncing orbits with positive and negative transverse directions share the same location. In this case, the trajectories in frequency-degenerate cavities at various degenerate states $|\Omega=P/Q\rangle$ are depicted in Fig.~\ref{spect}. For example at $|\Omega=1/4\rangle$ ($\phi_x=\pi/4$), positive and negative W-shaped trajectories coincide exactly, i.e. $|+\rangle=|u_0^\pm\rangle|u_1^\pm\rangle$ and $|-\rangle=|u_2^\pm\rangle|u_3^\pm\rangle$, as shown in Fig.~\ref{f1} in the main text. Additionally, in the four periodic round-trips, two share the first bouncing location and other two share the second one, i.e. $|1\rangle=|u_0^\pm\rangle|u_3^\pm\rangle$ and $|2\rangle=|u_1^\pm\rangle|u_2^\pm\rangle$. For the general spatial geometric mode, the ray equations for the 3-D periodic orbits can be written as:
\begin{equation}
\left\{
\begin{aligned}
u_{s}^{\pm }(z)&=\sqrt{{N_x}}\text{e}^{-\text{i}[{{\theta }_{s}}+{{\phi }_x}\pm {{\vartheta }}(z)]} \\
v_{s}^{\pm }(z)&=\sqrt{{N_y}}\text{e}^{-i[{{\theta }_{s}}+{{\phi }_y}\pm {{\vartheta }}(z)]}
\end{aligned}
\right.
\end{equation} 
where the dimensionless variable in the $y$-direction is similarly defined as $\widetilde{y}=\sqrt{2}y/w(z)$ with the ray equation $\widetilde{y}(z)=\operatorname{Re}[\sqrt{2}{v}_{s}^{\pm }(z)]$, here $+$ and $-$ in the symbol of $\pm$ indicate the positive and negative OAM states $|\pm\ell\rangle$, they together constitute a completed oscillation in cavity, as shown in Figs.~\ref{HO}(a) and (b) in the main text. For constituting a completed oscillation with both OAM states in a cavity, the phase factors yield $|\phi_x-\phi_y|=\pi/2$. 

The above is the ray representation of geometric modes in frequency-degenerate cavity. Hereinafter, we derive the wave representation coupled with the geometric modes and prove that it fulfills the SU(2) coherent state. The Gaussian wave packet with the central peak moving along the path $x=\operatorname{Re}(\sqrt{2}u)=\sqrt{2N}\cos (\omega t+{{\varphi }_{0}})$ can be derived as~\cite{rwd3}:
\begin{equation}
{{\pi}^{-1/4}}{{\text{e}}^{-{{x}^{2}}/2}}F(x,u)=\sum\limits_{n=0}^{\infty }{{{a}_{n}}\psi _{n}(x)}{{\text{e}}^{-\text{i}n(\omega t+{{\varphi }_{0}})}},\label{d7}
\end{equation}
where coefficients ${{a}_{n}}={N^{n/2}}{{e}^{-N/2}}/\sqrt{n!}=\sqrt{P(n,N)}$, where $P(n,N)$ is the Poisson distribution, $F(x,u)$ is the wave function of Schr{\"o}dinger coherent state and $\psi _{n}(x)$ 
is the Hermite--Gaussian function:
\begin{equation}
F(x,u)={{\operatorname{e}}^{-\left( {{u}^{2}}+|u{{|}^{2}}-2\sqrt{2}ux \right)/2}},
\end{equation}
\begin{equation}
{\psi }_{n}(x)={{\left( {{2}^{n}}n! \right)}^{-1/2}}{{\pi }^{-1/4}}{{H}_n}(x){{e}^{-{{x}^{2}}/2}}.
\end{equation}
The wave representation of a Gaussian wave packet moving along the $s$-th ray in a spatial geometric mode can be given by:
\begin{equation}
\Phi (\mathbf{r},u_{s}^{\pm },v_{s}^{\pm })=G(\mathbf{r})F(\widetilde{x},u_{s}^{\pm })F(\widetilde{y},v_{s}^{\pm }). \label{Phis}
\end{equation}
where $G(\mathbf{r})={{\pi }^{-1/2}}{\text{e}^{-({{{\widetilde{x}}}^{2}}+{{{\widetilde{y}}}^{2}})(1+\text{i}\widetilde{z})/2}}{\text{e}^{\mp\text{i}{{\vartheta}}(\widetilde{z})}}$ represents the fundamental mode Gaussian beam and $\mathbf{r}=(x,y,z)$ the Cartesian coordinates. In terms of Eq.~(\ref{Phis}), the resonant mode for the forward and backward components of a complete period is given by:
\begin{equation}
\Psi _{N_x,N_y}^{\pm }(\mathbf{r})=\frac{1}{Q}\sum\limits_{s=0}^{Q-1}{\Phi\left( \mathbf{r},u_{s}^{\pm },v_{s}^{\pm } \right){\text{e}^{\text{i}\left(N_x+N_y \right)\theta_s}}}, \label{nxny}
\end{equation}
where the phase term $\text{e}^{\text{i}\left( N_x+N_y\right)\theta_s}$ is associated
with the transverse frequency. For $N_x=0$ or $N_y=0$, Eq.~(\ref{nxny}) represents the planar geometric modes with ray structure on $(x,z)$ or $(y,z)$ plane; for $N_x=N_y\ne0$, circular vortex geometric modes; for $0\ne N_x\ne N_y\ne0$, elliptical vortex geometric modes. In the above description, $N_x$ or $N_y$ should be large enough to stimulate more ray-like properties, otherwise the pattern will be nearly a certain eigenmode. Hereinafter, we demonstrate the wave representation fulfills the form of SU(2) coherent state. Using planar trajectory $N_x=N$ and $N_y=0$ for convenience, it can be obtained that $v_{s}^{\pm }=0$ and $F(\widetilde y,0)=1$, and the planar geometric mode is given by:
\begin{equation}
\Psi _{N}^{\pm }(\mathbf{r})=\frac{1}{Q}\sum\limits_{s=0}^{Q-1}G(\mathbf{r}){F}\left( \widetilde{x},u_{s}^{\pm } \right){{\text{e}}^{\text{i}N{{\theta }_{s}}}},\label{d12}
\end{equation}
Substituting Eq.~(\ref{d7}) into Eq.~(\ref{d12}) and ignoring the constant coefficient, we get: 
\begin{align}
\nonumber
\Psi_{{N}}^{\pm}(\mathbf{r})&\propto\sum\limits_{s=0}^{Q-1}{G(\mathbf{r}){{\text{e}}^{\frac{{\widetilde{x}}^{2}}{2}}}\sum\limits_{n=0}^{\infty }{{{a}_{n}}{{\psi }_{n}}(\widetilde{x}){{\text{e}}^{-\text{i}n\left[ {{\theta }_{s}}+{{\phi }_{x}}\pm \vartheta (z) \right]}}}}{{\text{e}}^{\text{i}{{N}}{{\theta }_{s}}}}\\\nonumber
& =\sum\limits_{s=0}^{Q-1}{\sum\limits_{n=0}^{\infty }{{{a}_{n}}{{\psi }_{n}}(\widetilde{x}){{\text{e}}^{\frac{{{\widetilde{x}}^{2}}}{2}}}G(\mathbf{r}){{\text{e}}^{\mp \text{i}n\vartheta (z)-\text{i}n\left( {{\theta }_{s}}+{{\phi }_{x}} \right)+\text{i}N\theta_s}}}} \\ \nonumber
& \propto \sum\limits_{s=0}^{Q-1}{\sum\limits_{n=0}^{\infty }{{{a}_{n}}\psi _{n,0,l}^{(\text{HG})}(x,y,\pm z){{\text{e}}^{-\text{i}n{{\phi }_{x}}}}{{\text{e}}^{\text{i}\left( N-n \right){{\theta }_{s}}}}}} \\ 
& =\sum\limits_{n=0}^{\infty }{{{a}_{n}}\psi _{n,0,l}^{(\text{HG})}(x,y,\pm z){{\text{e}}^{-\text{i}n{{\phi }_{x}}}}\sum\limits_{s=0}^{Q-1}{{{\text{e}}^{\text{i}\left( N-n \right){{\theta }_{s}}}}}},\label{d13}
\end{align}
where the indices of HG modes should also fulfill the frequency-degenerate condition. Setting $n'=N-n$, the last term in the external summation notation of Eq.~(\ref{d13}) can be written as:
\begin{equation}
\sum\limits_{s=0}^{Q-1}{{{\text{e}}^{\text{i}n'{{\theta }_{s}}}}}=\sum\limits_{s=0}^{Q-1}\text{e}^{\text{i}n'\frac{P}{Q}2\pi s}. \label{d14}
\end{equation}
When $n'=KQ$ ($K\in\mathbb{Z}$), $\text{e}^{\text{i}n'\frac{P}{Q}2\pi s}=\text{e}^{\text{i}2\pi KPs}=1$ and Eq.~(\ref{d14}) is equal to a constant $Q$; when $n'\ne KQ$, Eq.~(\ref{d14}) is always a sum of the complex numbers uniformly distributed on the unit circle of the complex plane, thus it should be zero. Then Eq.~(\ref{d13}) can be reduced as:
\begin{align}
\nonumber
\Psi_{{N}}^{\pm}(\mathbf{r})&\propto\sum\limits_{K=0}^{N}{{{a}_{n_0+KQ}}{\text{e}}^{-\text{i}KQ\phi _x}\psi _{n_0+KQ,0,l_0-KP}^{(\text{HG})}(x,y,\pm z)} \\\nonumber
&= \sum\limits_{K=0}^{N}\sqrt{P(n_0+KQ,N)}{\text{e}}^{-\text{i}KQ\phi _x}|K,N\rangle \\ \nonumber
&\propto\sum\limits_{K=0}^{N}\sqrt{B(K;N,\frac{1}{2})}{\text{e}}^{-\text{i}KQ\phi _x}|K,N\rangle \\
&=\frac{\text{1}}{{{\text{2}}^{N/2}}}\sum\limits_{K=0}^{N}{{{\left( \begin{matrix}
			N  \\
			K  \\
			\end{matrix} \right)}^{1/2}}{\text{e}}^{\text{i}K\phi}|K,N\rangle},\label{d15}
\end{align}
where the coherent state phase $\phi=Q\phi_x$, and the Poisson distribution $P(n_0+KQ,N)$ is approximated by Binomial distribution $B(K;N,1/2)$ when $N=4Q^2n_0$ is large enough, where $B(k;n,p)=\left( \begin{matrix}
n  \\
k  \\
\end{matrix} \right){{p}^{k}}{{\left( 1-p \right)}^{n-k}}$ is Binomial distribution, according to the central-limit theorem. Then the laser mode Eq.~(\ref{d15}) shares the same form of SU(2) coherent state as Eq.~(\ref{phi}).

Therefore, the SU(2) wave-packet in frequency-degenerate cavity has the property of ray-wave duality, the laser mode can be not only characterized by the wave function representation but also coupled with classical oscillating trajectory given by ray representation. The preponderance of wave-like or ray-like property can be actually controlled by the parameters in SU(2) wave-packet.

\subsection{Generation of SU(2) geometric beams}\label{AE}
In section~\ref{AC}, we demonstrated that the frequency-degenerate cavity fulfills the condition of generating SU(2) wave-packet. Hereinafter, we will demonstrate the actual method to generate SU(2) planar geometric beams in a frequency-degenerate cavity by off-axis pumping. Considering the gain distribution $f(x,y,z)$ , the resonant modes of the laser system pumped by a localized source can be solved from the inhomogeneous Helmholtz equation: 
\begin{equation}
\left( {{\nabla }^{2}}+{{\widetilde{k}}^{\text{2}}} \right)\Psi (x,y,z)={{\eta }_{c}}f(x,y,z),
\label{nwe}
\end{equation}
where $\widetilde{k}=k_0+\text{i}\alpha$, $k_0=2\pi /\lambda_0 $ is the given wave number of the laser with given wavelength $\lambda_0$, the factor $\eta_c$ represents the conversion efficiency for the excitation source, and $\alpha$ is a small loss parameter including losses from the scattering, the absorption, and the output coupling. Based on the orthogonality and completeness of HG modes, the lasing mode and the source distribution can be expressed as the superposition of HG eigenmodes:
\begin{equation}
\Psi(x,y,z)=\sum\limits_{n,m,l}{{{a}_{n,m,l}}\psi _{n,m,l}^{(\text{HG})}(x,y,z)}
\label{phihg}
\end{equation}
and
\begin{equation}
\eta_{c}F(x,y,z)=\sum\limits_{n,m,l}{{b_{n,m,l}}\psi _{n,m,l}^{(\text{HG})}(x,y,z)}
\label{fhg}
\end{equation}
Substituting Eqs.~(\ref{phihg}) and (\ref{fhg}) into the wave equation Eq.~(\ref{nwe}), the relationship
between the coefficients $a_{n,m,l}$ and $b_{n,m,l}$ can be found:
\begin{equation}
{{a}_{n,m,l}}={{b}_{n,m,l}}/({{\widetilde{k}}^{2}}-k_{n,m,l}^{2}).
\label{ab}
\end{equation}
Considering the condition of $\alpha\ll k_0$ and substituting Eq.~(\ref{ab}) into Eq.~(\ref{phihg}), the eigenmode expansion of the resonant mode is given by:
\begin{equation}
\Psi(x,y,z)=\sum\limits_{n,m,l}{\frac{{{b}_{n,m,l}}}{({{k}^{2}_0}-k_{n,m,l}^{2})+2\text{i}\alpha k_0}}\psi _{n,m,l}^{(\text{HG})}(x,y,z).\label{Psi}
\end{equation}
With the orthonormal property of eigenmodes, the coefficient $b_{n,m,l}$ is given by:
\begin{equation}
{{b}_{n,m,l}}={{\eta }_{c}}\iiint{\psi _{n,m,l}^{(\text{HG})}(x,y,z)f(x,y,z)\text{d}x\text{d}y\text{d}z}.
\label{b}
\end{equation}
Supposing the pump light is a Gaussian beam along $z$-axis with focused waist at $z=z_c$ position and an off-axis displacement $\Delta x$ at $x$-direction, which is larger than pump spot size, $\Delta x\gg w_0$, also supposing that the gain medium center is located at the pump waist with the thickness much less than the cavity length, i.e. $L_c\ll L$, the pump source distribution can be written as:
\begin{equation}
f(x,y,z)=\frac{2}{\pi {{w}^{2}}(z-z_c){{L}_{c}}}{{\operatorname{e}}^{-\frac{{{(x-\Delta x)}^{2}}+{{y}^{2}}}{{{w}^{2}}(z-z_c)}}},
\label{G}
\end{equation}
for $|z-z_c|\le L_c/2$, here $w(z)$ represents the pump beam radius distribution. Since the longitudinal distribution of the pump source is nearly uniform, the coefficient $b_{n,m,l}$ related to the source term $f(x,y,z)$ can be considered to be independent of the index $l$ in the neighborhood of the central index $l_0$.
Consequently, the integral in Eq.~(\ref{b}) can be approximately reduced as:
\begin{equation}
{{b}_{n,m,l}}=\frac{2\eta }{\pi w_{0}^{2}{{L}_{c}}}\iint{\psi _{n,m,l}^{(\text{HG})}(x,y,0)}{{\operatorname{e}}^{-\frac{{{(x-\Delta x)}^{2}}+{{y}^{2}}}{w_{0}^{2}}}}\text{d}x\text{d}y,\label{bnm}
\end{equation}
where $\eta$ is a constant that includes the effective conversion efficiency $\eta_{c}$ and the overlap integral in the longitudinal
direction, here $w_0$ represents the pump beam radius at its waist. Substituting Eqs.~(\ref{hg}) and (\ref{G}) into Eq.~(\ref{bnm}), setting $X=\sqrt{2}x/{{w}_{0}}$, $Y=\sqrt{2}y/{{w}_{0}}$, ${{n}_{x}}={{\left( \Delta x/{{w}_{0}} \right)}^{2}}$, and applying generating function of the Hermite polynomials, $b_{n,m,l}$ can be derived as:
\begin{widetext}
	\begin{align}
	\nonumber
	{{b}_{n,m,l}}&=\frac{2\eta }{\pi w_{0}^{2}{{L}_{c}}}\int_{-\infty}^{\infty}\int_{-\infty}^{\infty}{\frac{1}{\sqrt{{{2}^{m+n-1}}\pi m!n!}}\frac{1}{{{w}_{0}}}{{H}_{n}}\left( \frac{\sqrt{2}x}{{{w}_{0}}} \right){{H}_{m}}\left( \frac{\sqrt{2}y}{{{w}_{0}}} \right){{\operatorname{e}}^{-\frac{{{x}^{2}}+{{y}^{2}}}{w_{0}^{2}}}}{{\operatorname{e}}^{-\frac{{{\left( x-\Delta x \right)}^{2}}+{{y}^{2}}}{w_{0}^{2}}}}\text{d}x\text{d}y} \\\nonumber 
	& =\frac{\eta }{\pi {{w}_{0}}{{L}_{c}}}\int_{-\infty}^{\infty}\int_{-\infty}^{\infty}{\frac{1}{\sqrt{{{2}^{m+n-1}}\pi m!n!}}{{H}_{n}}\left( X \right){{H}_{m}}\left( Y \right){{\operatorname{e}}^{-\frac{{{X}^{2}}+{{Y}^{2}}}{2}}}{{\operatorname{e}}^{-{{\left( \frac{X}{\sqrt{2}}-{{n}_{x}} \right)}^{2}}-{{\left( \frac{Y}{\sqrt{2}} \right)}^{2}}}}\text{d}X\text{d}Y} \\\nonumber 
	& =\frac{\eta }{\pi {{w}_{0}}{{L}_{c}}}{{\delta }_{m,0}}\int_{-\infty}^{\infty}{\frac{1}{\sqrt{{{2}^{n-1}}\pi n!}}{{H}_{n}}\left( X \right){{\operatorname{e}}^{-\frac{{{X}^{2}}}{2}}}{{\operatorname{e}}^{-\frac{{{X}^{2}}}{2}+\sqrt{2{{n}_{x}}}X-{{n}_{x}}}}\text{d}X} \\ \nonumber
	& =\frac{\eta }{\pi {{w}_{0}}{{L}_{c}}}\frac{1}{\sqrt{{{2}^{n-1}}\pi n!}}{{\operatorname{e}}^{-\frac{{{n}_{x}}}{2}}}{{\delta }_{m,0}}\int_{-\infty}^{\infty}{{{H}_{n}}\left( X \right){{\operatorname{e}}^{-{{X}^{2}}}}{{\operatorname{e}}^{\sqrt{2{{n}_{x}}}X-\frac{{{n}_{x}}}{2}}}\text{d}X} \\ \nonumber
	& =\frac{\eta }{\pi {{w}_{0}}{{L}_{c}}}\frac{1}{\sqrt{{{2}^{n-1}}\pi n!}}{{\operatorname{e}}^{-\frac{{{n}_{x}}}{2}}}{{\delta }_{m,0}}\int_{-\infty}^{\infty}{{{H}_{n}}\left( X \right){{\operatorname{e}}^{-{{X}^{2}}}}\left[ \sum\limits_{p=0}^{\infty }{\frac{{{H}_{p}}\left( X \right){{\left( \sqrt{\frac{{n}_{x}}{2}} \right)}^{p}}}{p!}} \right]\text{d}X} \\ \nonumber
	& =\frac{\eta }{\pi {{w}_{0}}{{L}_{c}}}\frac{1}{\sqrt{{{2}^{n-1}}\pi n!}}\frac{{{\left( {{n}_{x}} \right)}^{\frac{n}{2}}}}{\sqrt{{{2}^{n}}}}\frac{1}{n!}{{\operatorname{e}}^{-\frac{{{n}_{x}}}{2}}}{{\delta }_{m,0}}\int_{-\infty}^{\infty}{{{H}_{n}}\left( X \right){{H}_{n}}\left( X \right){{\operatorname{e}}^{-{{X}^{2}}}}\text{d}X} \\ \nonumber
	& =\frac{\eta {{\delta }_{m,0}}}{\pi {{w}_{0}}{{L}_{c}}}\frac{1}{\sqrt{{{2}^{n-1}}\pi n!}}\frac{{{\left( {{n}_{x}} \right)}^{\frac{n}{2}}}}{\sqrt{{{2}^{n}}}}\frac{1}{n!}{{\operatorname{e}}^{-\frac{{{n}_{x}}}{2}}}{{\delta }_{m,0}}\left( \sqrt{\pi }{{2}^{n}}n! \right) \\ 
	& =\frac{\sqrt{2}\eta }{\pi {{w}_{0}}{{L}_{c}}}\frac{{{\left( {{n}_{x}} \right)}^{\frac{n}{2}}}}{\sqrt{n!}}{{\operatorname{e}}^{-\frac{{{n}_{x}}}{2}}}{{\delta }_{m,0}}=\frac{\sqrt{2}\eta }{\pi {{w}_{0}}{{L}_{c}}}\sqrt{P\left(n,{{n}_{x}}\right)}{{\delta }_{m,0}},\label{bp}
	\end{align}
	where $\delta_{i,j}$ is the Kronecker delta. Because $\Delta x\gg w_0$ and similarly $n_x\gg\sqrt{2}$, there should be no overlap between the pump spot and the high-order mode at $y$-direction, we can use $\delta_{m,0}$ to simplify the formula. We also use the orthogonality with respect to the weight function of Hermite polynomials, $\int_{-\infty}^{\infty}H_m(x)H_n(x)\text{e}^{-x^2}\text{d}x=\sqrt{\pi}2^nn!\delta_{n,m}$, to reduce the integral in Eq.~(\ref{bp}). Based on Eq.~(\ref{f}) and relationship ${{k}_{n,m,l}}=2\pi {{f}_{n,m,l}}/c$, and no high-order mode at $y$-axis here, the expressions of center and variable wave numbers are derived as ${{k}_{0}}=\pi \left[ {{l}_{0}}+({{n}_{0}}+1)\Omega  \right]/L$, and ${{k}_{n,m,l}}=\pi \left[ {{l}}+({{n}}+1)\Omega  \right]/L$. Substituting them and Eq.~(\ref{bp}) into Eq.~(\ref{Psi}), we can derive the expression of laser mode:
	\begin{align}
	\nonumber
	\Psi (x,y,z)&=\frac{\sqrt{2}\eta }{\pi {{w}_{0}}{{L}_{c}}}{{\delta }_{m,0}}\sum\limits_{n,m,l}{\frac{\sqrt{P({n,n_x})}}{(k+{{k}_{n,m,l}})(k-{{k}_{n,m,l}})+2\text{i}\alpha k}}\psi _{n,m,l}^{(\text{HG})}(x,y,z) \\\nonumber 
	& =\frac{\sqrt{2}\eta }{\pi {{w}_{0}}{{L}_{c}}}\sum\limits_{n}{\sum\limits_{l}{\frac{\sqrt{P({n,n_x})}}{{{\pi }^{2}}/{{L}^{2}}\{[l+{{l}_{0}}+(n+{{n}_{0}}+2)\Omega ][{{l}_{0}}-l+({{n}_{0}}-n)\Omega ]+2\text{i}\gamma [{{l}_{0}}+({{n}_{0}}+1)\Omega ]\}}}\psi _{n,0,l}^{(\text{HG})}(x,y,z)} \\ \nonumber
	& =\frac{\sqrt{2}\eta }{\pi {{w}_{0}}{{L}_{c}}}\frac{{{L}^{2}}}{{{\pi }^{2}}}\sum\limits_{n}{\sum\limits_{l}{\frac{\sqrt{P({n,n_x})}}{[l+{{l}_{0}}+(n+{{n}_{0}}+2)\Omega ][{{l}_{0}}-l+({{n}_{0}}-n)\Omega ]+2\text{i}\gamma [{{l}_{0}}+({{n}_{0}}+1)\Omega ]}}\psi _{n,0,l}^{(\text{HG})}(x,y,z)} \\
	& =\frac{\sqrt{2}\eta{{L}^{2}}}{{{\pi }^{3}}{{w}_{0}}{{L}_{c}}}\sum\limits_{n}{\sum\limits_{l}{\frac{\sqrt{P({n,n_x})}}{\beta [({{l}_{0}}-l)+({{n}_{0}}-n)\Omega ]+\text{i}\gamma }}\psi _{n,0,l}^{(\text{HG})}(x,y,z)}, \label{PsiP} 
	\end{align}
	where $\gamma =\alpha L/\pi $ and $\beta =\frac{l+{{l}_{0}}+(n+{{n}_{0}}+2)\Omega }{2[{{l}_{0}}+({{n}_{0}}+1)\Omega ]}=\frac{k+{{k}_{n,m,l}}}{2k}\approx 1$. Note that the value of the parameter $n_x$ signifies the magnitude of the off-axis displacement, also indicates that the maximum contribution in the resonant mode comes from the eigenmode with the transverse index $n$ to be closest to the value $n_0$. Therefore, we can take the parameter $n_x=n_0$ for convenience. For a Poisson distribution with large parameter, it can be approximated using a Gaussian distribution of an effective range near its mean, then the effective range of mode index $n$ can be limited as $|n-n_0|\le n'$ ($n'=2\sqrt{{{n}_{0}}}$) for the sum in Eq.~(\ref{PsiP}):
	\begin{equation}
	\Psi (x,y,z)=\frac{\sqrt{2}\eta {{L}^{2}}}{{{\pi }^{3}}{{w}_{0}}{{L}_{c}}}\sum\limits_{l={{l}_{0}}-l'}^{{{l}_{0}}+l'}{\sum\limits_{n={{n}_{0}}-n'}^{{{n}_{0}}+n'}{\frac{\sqrt{P({n,n_0})}}{\beta [({{l}_{0}}-l)+({{n}_{0}}-n)\Omega ]+\text{i}\gamma }}}\psi _{n,m,l}^{(\text{HG})}(x,y,z). \label{Psin}
	\end{equation}
	Considering $\gamma =\alpha L/\pi $ where $\alpha \sim {{10}^{-6}}$m and $\Omega =P/Q$ is a simple fraction number, when $\left| ({{l}_{0}}-l)+({{n}_{0}}-n)\Omega  \right|\ne 0$, there should be $\left| ({{l}_{0}}-l)+({{n}_{0}}-n)\Omega  \right|\gg \left| \text{i}\gamma  \right|$, so that we can neglect the terms of corresponding eigenstates and only consider the HG modes that satisfy $\left| ({{l}_{0}}-l)+({{n}_{0}}-n)\Omega  \right|=0$, which can also be written as $-\left( {{l}_{0}}-l \right)/\left( {{n}_{0}}-n \right)=\Omega =P/Q$. Defining $K=(l-l_0)/P=-(n_0-n)/Q$, Eq.~(\ref{Psin}) can be simplified with only one accumulative symbol:
	\begin{align}
	\Psi(x,y,z)=\frac{\sqrt{2}\eta {{L}^{2}}}{{{\pi }^{3}}{{w}_{0}}{{L}_{c}}}\sum\limits_{K=-n'/Q}^{n'/Q}{\frac{\sqrt{P({n_0-KQ,n_0})}}{\text{i}\alpha L/\pi }}\psi _{{{n}_{0}}+KQ,0,{{l}_{0}}-KP}^{(\text{HG})}(x,y,z).\label{e12}
	\end{align}
\end{widetext}
Setting $\bar{n}_0=n_0-n'$, $\bar{l}_0=l_0-l'$, and $N=4Q^2n_0$, and according to the central-limit theorem in maximum likelihood estimation, the Poission distribution $P(n_0-KQ,n_0)$ can be approximated by Binomial distribution $B(K;N,1/2)$ when $N$ is large enough. Then the laser mode Eq.~(\ref{e12}) can be derived into the form of SU(2) coherent state:
\begin{align}
\nonumber
\Psi (x,y,z)&\propto \sum\limits_{K=0}^{N}{\sqrt{B(K;N,\frac{1}{2})}\psi _{{\bar{n}_{0}}+KQ\,,0,\,{\bar{l}_{0}}-KP}^{(\text{HG})}}(x,y,z) \\ 
&=\frac{\text{1}}{{{\text{2}}^{N/2}}}\sum\limits_{K=0}^{N}{{{\left( \begin{matrix}
			N  \\
			K  \\
			\end{matrix} \right)}^{1/2}}\psi _{{\bar{n}_{0}}+KQ,0,\bar{l}_0-PK}^{(\text{HG})}(x,y,z)}, 
\end{align}
sharing the same form of Eq.~(\ref{phi}). If we stretch the pump size along $y$-axis to stimulate the high-order mode at $y$-direction and control the astigmatism in pumping light to lock additive phase among eigenmodes, the laser mode can be controlled as more general SU(2) coherent state as Eq.~(\ref{pgm}). The laser modes can also be transformed into vortex geometric modes Eq.~(\ref{vgm}) by astigmatic mode converter that transforms the HG bases into corresponding LG bases. Hereto, we have proved that the frequency-degenerate cavity with off-axis pumping can emit the laser mode of SU(2) coherent state.

\begin{figure*}
	\centering
	\includegraphics[width=\linewidth]{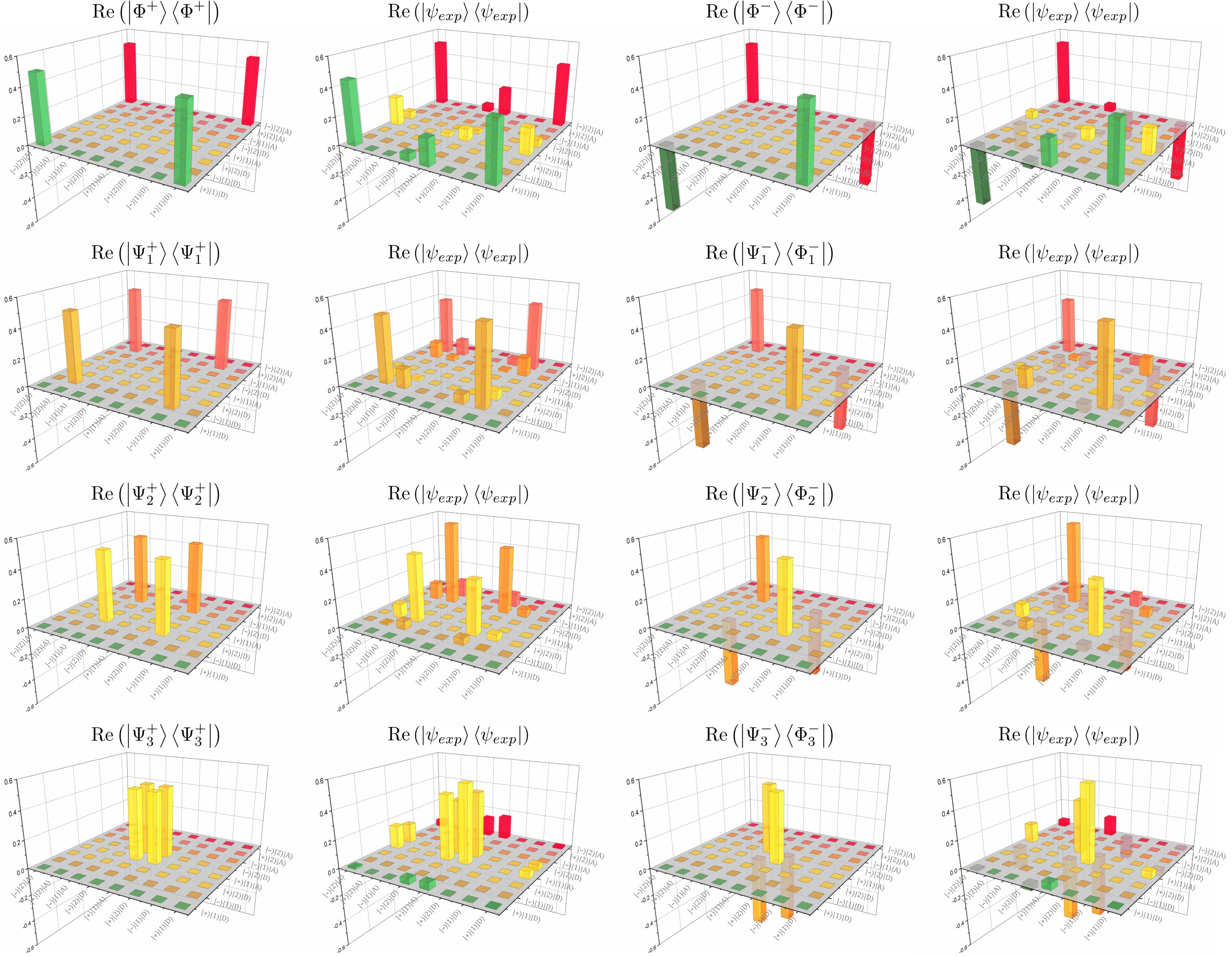}
	\caption{\footnotesize \textbf{State tomography.} The theoretical and experimental results of the tomography of density matrices for various GHZ states .} 
	\label{tomo}
\end{figure*}

\subsection{General SU(2) vector beams}\label{AF}
By elaborate external modulation, a normal SU(2) geometric beam can be modulated as a general SU(2) vector beam, where arbitrary amplitude, phase, and polarization for each orbit can be modulated. By modifying Eq.~(\ref{nxny}), a general SU(2) vector beam can be given by:
\begin{equation}
\mathbf{\Psi }_{N_x,N_y}^{\pm }(\mathbf{r})=\frac{1}{Q}\sum\limits_{s=0}^{Q-1}A_s\text{e}^{\phi_s}\mathbf{J}_s\Phi_{N_x,N_y}^{(s)}\left( \mathbf{r} \right), \label{gv}
\end{equation}
where we set $\Phi_{N_x,N_y}^{(s)}=\Phi\left( \mathbf{r},u_{s}^{\pm },v_{s}^{\pm } \right){\text{e}}^{\text{i}\left(N_x+N_y \right)\theta_s}$ for convenience; $A_s$, $\phi_s$, and $\mathbf{J}_s$ are the amplitude, phase, and polarization Jones vector of light at the $s$-th orbit. In degenerate state $|\Omega=1/4\rangle$, we can firstly control the SU(2) beam into ray-like state and then make special modulations to realize various classical GHZ states [see Eqs.~(\ref{cghz1}) to (\ref{cghz4}) in the next section~\ref{AG}] as specific cases: 
\begin{itemize}	
	\item
	$|{\Phi }^{+}\rangle$: $A_0=A_2=\frac{1}{\sqrt{2}}$, $A_1=A_3=0$, $\phi_0=\phi_2$, $\mathbf{J}_0=|D\rangle$ and $\mathbf{J}_2=|A\rangle$;
	\item
	$|{\Phi }^{-}\rangle$: $A_0=A_2=\frac{1}{\sqrt{2}}$, $A_1=A_3=0$, $\phi_0=\phi_2+\pi$, $\mathbf{J}_0=|D\rangle$ and $\mathbf{J}_2=|A\rangle$; 
	\item
	$|\Psi_1^+\rangle$: $A_1=A_3=\frac{1}{\sqrt{2}}$, $A_0=A_2=0$, $\phi_1=\phi_3$, $\mathbf{J}_1=|D\rangle$ and $\mathbf{J}_3=|A\rangle$; 	
	\item
	$|\Psi_1^-\rangle$: $A_1=A_3=\frac{1}{\sqrt{2}}$, $A_0=A_2=0$, $\phi_1=\phi_3+\pi$, $\mathbf{J}_1=|D\rangle$ and $\mathbf{J}_3=|A\rangle$; 
	\item
	$|\Psi_2^+\rangle$: $A_0=A_3=\frac{1}{\sqrt{2}}$, $A_1=A_2=0$, $\phi_0=\phi_3$, $\mathbf{J}_0=|D\rangle$ and $\mathbf{J}_3=|A\rangle$; 	
	\item
	$|\Psi_2^-\rangle$: $A_0=A_3=\frac{1}{\sqrt{2}}$, $A_1=A_2=0$, $\phi_0=\phi_3+\pi$, $\mathbf{J}_0=|D\rangle$ and $\mathbf{J}_3=|A\rangle$; 
	\item
	$|\Psi_3^+\rangle$: $A_0=A_2=\frac{1}{\sqrt{2}}$, $A_1=A_3=0$, $\phi_0=\phi_2$, $\mathbf{J}_0=|A\rangle$ and $\mathbf{J}_2=|D\rangle$; 	
	\item
	$|\Psi_3^-\rangle$: $A_0=A_2=\frac{1}{\sqrt{2}}$, $A_1=A_3=0$, $\phi_0=\phi_2+\pi$, $\mathbf{J}_0=|A\rangle$ and $\mathbf{J}_2=|D\rangle$;		
\end{itemize}

\begin{figure}
	\centering
	\includegraphics[width=\linewidth]{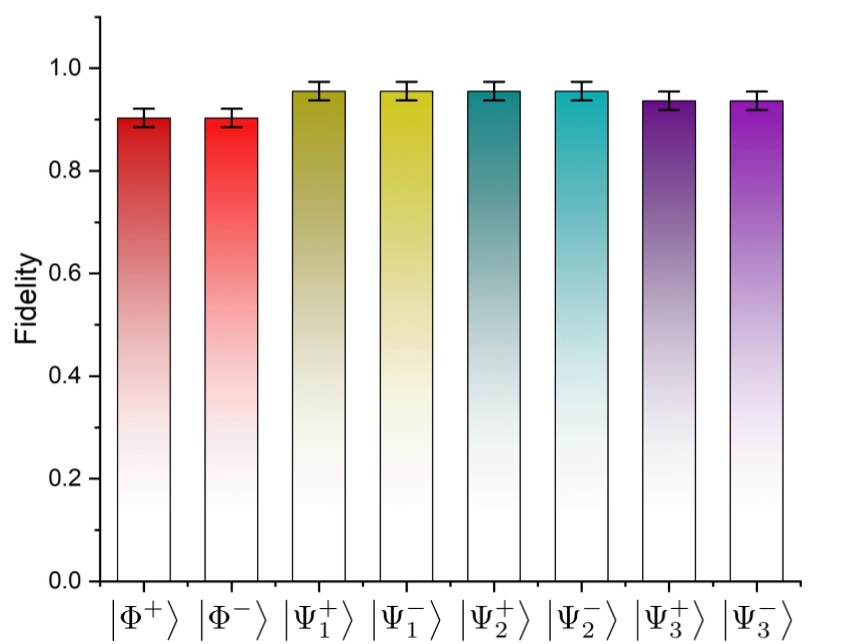}
	\caption{\footnotesize \textbf{Fidelity.} The experimental results of fidelities for various GHZ states.} 
	\label{fid}
\end{figure}

When an SU(2) vector beam was controlled into ray-like state with large enough $N_x$ and $N_y$ without interference among lights on sub-orbits, the wavefunction along different orbits are independent without interference to each other, thus intensity pattern can be simplified as:
\begin{equation}
\left\|\mathbf{\Psi }_{N_x,N_y}^{\pm }(\mathbf{r})\right\|^2=\frac{1}{Q^2}\sum\limits_{s=0}^{Q-1}A^2_s\left|\Phi_{N_x,N_y}^{(s)}\left( \mathbf{r} \right)\right|^2. \label{I}
\end{equation}
where $\left\|\mathbf{\Psi }\right\|$ is the Frobenius norm of vector $\mathbf{\Psi }$. Hereinafter, we derive the expression for the polarization projection states. The Jones matrix for a linear polarizer with a inclined angle of $\theta_P$ is ${{\mathbf{J}}_{P}}=\left[ \begin{matrix}
\cos {{\theta }_{P}} & 0  \\
0 & \sin {{\theta }_{P}}  \\
\end{matrix} \right]$, which can project the light into the linear polarization state with inclined angle of $\theta_P$. the SU(2) vector beam after projection can be given by:
\begin{equation}
\widetilde{\mathbf{\Psi }}_{N_x,N_y}^{\pm }(\mathbf{r})=\frac{1}{Q}\sum\limits_{s=0}^{Q-1}A_s\text{e}^{\phi_s}\mathbf{J}_P\mathbf{J}_s\Phi_{N_x,N_y}^{(s)}\left( \mathbf{r} \right).
\end{equation}
When $N_x$ and $N_y$ are both large enough, the lights on various orbits cannot make interference to each other, the intensity pattern can be given by.
\begin{equation}
\left\|\widetilde{\mathbf{\Psi }}_{N_x,N_y}^{\pm }(\mathbf{r})\right\|^2=\frac{1}{Q^2}\sum\limits_{s=0}^{Q-1}A^2_s\left\|\mathbf{J}_P\mathbf{J}_s \right\|^2\left|\Phi_{N_x,N_y}^{(s)}\left( \mathbf{r} \right)\right|^2, 
\end{equation}
The 3-D intensity patterns of various GHZ states with various polarization projection states ($\theta_P=0$, $\pi/2$, $\pi$, and $3\pi/2$) are shown in Fig.~(\ref{P}), where we set $N_x=N_y$ in simulation. When the vortex geometric mode is reduced into planar geometric mode, i.e. $N_x$ or $N_y=0$, we can directly use $\langle\Psi_{N_x,N_y}^\pm|\Psi_{N_x,N_y}^\pm\rangle$ to observe the interference effect among sub-orbits, as shown in Fig.~{\ref{Bell}} for various GHZ states. By experimentally measuring the amplitude and polarization of each orbit, we can evaluate the experimental state:
\begin{align}
\nonumber
\left| \psi_{exp}  \right\rangle =& \text{ }{\alpha_1}\left| + \right\rangle \left| 1 \right\rangle \left| D \right\rangle +{{\alpha }_{2}}\left| - \right\rangle \left| 1 \right\rangle \left| D \right\rangle +{{\alpha }_{3}}\left| + \right\rangle \left| 2 \right\rangle \left|D \right\rangle  \\ \nonumber
&+{{\alpha }_{4}}\left| + \right\rangle \left| 1 \right\rangle \left| A\right\rangle +{{\alpha }_{5}}\left| - \right\rangle \left| 2 \right\rangle \left| D \right\rangle +{{\alpha }_{6}}\left| - \right\rangle \left| 1 \right\rangle \left|A \right\rangle  \\ 
&+{{\alpha }_{7}}\left|+ \right\rangle \left| 2 \right\rangle \left|A \right\rangle +{{\alpha }_{8}}\left|- \right\rangle \left|2 \right\rangle \left|A \right\rangle,
\end{align}
reconstitute the density matrix $\hat{\rho}_{exp}=|\psi_{exp}\rangle\langle\psi_{exp}|$ for each GHZ state, and calculate the fidelities $F=\langle\Phi^\pm|\hat{\rho}_{exp}|\Phi^\pm\rangle$ and $F=\langle\Psi^\pm_i|\hat{\rho}_{exp}|\Psi^\pm_i\rangle$ ($i=1,2,3$) comparing with the theoretical density matrices of GHZ states. The theoretical and experimental tomography of density matrices and fidelities for various GHZ states are shown in Fig.~\ref{tomo} and Fig.~\ref{fid}.

\subsection{Classical GHZ states}\label{AG}
Maximally entangled states of two particles are called Bell states, acting as 4 eigenstates of 4-D Hilbert space, and those of $N$ ($N\ge3$) particles are called GHZ states, acting as $2^N$ eigenstates of $2^N$-D Hilbert space~\cite{pan1998greenberger}. The 4 Bell states in 2 maximumly entangled groups for polarization-entangled photon pair are given by:
\begin{equation}
\left| {{\phi }^{\pm }} \right\rangle =\frac{\left| H \right\rangle \left| V \right\rangle \pm \left| V \right\rangle \left| H \right\rangle }{\sqrt{2}},
\label{b1}
\end{equation}
\begin{equation}
\left| {{\psi }^{\pm }} \right\rangle =\frac{\left| H \right\rangle \left| H \right\rangle \pm \left| V \right\rangle \left| V \right\rangle }{\sqrt{2}}.
\label{b2}
\end{equation}
They can linearly express a general 4-D polarization-entangled state:
\begin{equation}
\left| \psi  \right\rangle =\alpha \left| H \right\rangle \left| H \right\rangle +\beta \left| H \right\rangle \left| V \right\rangle +\gamma \left| V \right\rangle \left| V \right\rangle +\delta \left| V \right\rangle \left| V \right\rangle.
\label{g4d}
\end{equation}
The 8 GHZ states in 4 maximumly entangled groups for three-photon polarization-entanglement are given by~\cite{pan1998greenberger}:
\begin{equation}
\left| {{\Phi }^{\pm }} \right\rangle =\frac{\left| H \right\rangle \left| H \right\rangle \left| H \right\rangle \pm \left| V \right\rangle \left| V \right\rangle \left| V \right\rangle }{\sqrt{2}},
\label{ghz1}
\end{equation}
\begin{equation}
\left| \Psi _{1}^{\pm } \right\rangle =\frac{\left| V \right\rangle \left| H \right\rangle \left| H \right\rangle \pm \left| H \right\rangle \left| V \right\rangle \left| V \right\rangle }{\sqrt{2}},
\label{ghz2}
\end{equation}
\begin{equation}
\left| \Psi _{2}^{\pm } \right\rangle =\frac{\left| H \right\rangle \left| V \right\rangle \left| H \right\rangle \pm \left| V \right\rangle \left| H \right\rangle \left| V \right\rangle }{\sqrt{2}},
\label{ghz3}
\end{equation}
\begin{equation}
\left| \Psi _{3}^{\pm } \right\rangle =\frac{\left| H \right\rangle \left| H \right\rangle \left| V \right\rangle \pm \left| V \right\rangle \left| V \right\rangle \left| H \right\rangle }{\sqrt{2}}.
\label{ghz4}
\end{equation}
They can linearly express a general 8-D polarization-entangled state:
\begin{align}
	\nonumber
	\left| \psi  \right\rangle =& \text{ }{\alpha_1}\left| H \right\rangle \left| H \right\rangle \left| H \right\rangle +{{\alpha }_{2}}\left| V \right\rangle \left| H \right\rangle \left| H \right\rangle +{{\alpha }_{3}}\left| H \right\rangle \left| V \right\rangle \left| H \right\rangle  \\ \nonumber
	&+{{\alpha }_{4}}\left| H \right\rangle \left| H \right\rangle \left| V \right\rangle +{{\alpha }_{5}}\left| V \right\rangle \left| V \right\rangle \left| H \right\rangle +{{\alpha }_{6}}\left| V \right\rangle \left| H \right\rangle \left| V \right\rangle  \\ 
	&+{{\alpha }_{7}}\left| H \right\rangle \left| V \right\rangle \left| V \right\rangle +{{\alpha }_{8}}\left| V \right\rangle \left| V \right\rangle \left| V \right\rangle.
\label{g8d} 
\end{align}
By comparing the representation of vector beam ($|\psi\rangle=|\ell_1\rangle|H\rangle + |\ell_2\rangle|V\rangle$) and Eqs.~(\ref{b1}) and (\ref{b2}), the vector beams show an excellent classical picture corresponding to the quantum Bell states, because the non-separability of transverse mode and polarization DoFs in vector beam is just corresponding to the non-separability of two polarization states in entangled photon pair. Thus this exotic non-separability in vector beams is also called classical entanglement. By manipulating the spatial modes $|u_1\rangle$ and $|u_2\rangle$, a completed qubit can be revealed by specially designed vector beams such as the higher-order Poincar\'e sphere~\cite{hps2} and the hybrid-order Poincar\'e sphere~\cite{hyps2}. Moreover, quantum measure and tomography methods can be referred to measure degree of non-separability in classical entanglement~\cite{mclaren2015measuring}. However, this previous works are constrained within the 2-D Bell states. In order to transfer GHZ states into classical entanglement, we should find more DoFs beyond transverse mode and polarization. It seems difficult to find more DoFs because the transverse mode and polarization are the basic elements constituting a common laser beam. In contrast to the common beams, the geometric beams in SU(2) coherent states have a coupling effect between the transverse and longitudinal modes induced by the frequency degeneracy where the spatial patterns vary with the longitudinal distance~\cite{su21}. In an SU(2) coherent state, the actual 3-D wave-packet pattern can be located at some periodic oscillating orbits, manifesting the ray-wave duality~\cite{rwd1,rwd2}. We can elaborately modulate the intensity and polarization of each periodic orbit to manipulate three DoFs and realize GHZ states alternatively: the first DoF is the oscillating direction, the second is the location of periodic orbits, the third is the polarization, then the classical GHZ states yield:
\begin{equation}
\left| {{\Phi }^{\pm }} \right\rangle =\frac{\left| + \right\rangle \left| 1 \right\rangle \left| D \right\rangle \pm \left| - \right\rangle \left| 2 \right\rangle \left| A \right\rangle }{\sqrt{2}},
\label{cghz1}
\end{equation}
\begin{equation}
\left| \Psi _{1}^{\pm } \right\rangle =\frac{\left| - \right\rangle \left| 1 \right\rangle \left| D \right\rangle \pm \left| + \right\rangle \left| 2 \right\rangle \left| A \right\rangle }{\sqrt{2}},
\label{cghz2}
\end{equation}
\begin{equation}
\left| \Psi _{2}^{\pm } \right\rangle =\frac{\left| + \right\rangle \left| 2 \right\rangle \left| D \right\rangle \pm \left| - \right\rangle \left| 1 \right\rangle \left| A \right\rangle }{\sqrt{2}},
\label{cghz3}
\end{equation}
\begin{equation}
\left| \Psi _{3}^{\pm } \right\rangle =\frac{\left| + \right\rangle \left| 1 \right\rangle \left| A \right\rangle \pm \left| - \right\rangle \left| 2 \right\rangle \left| D \right\rangle }{\sqrt{2}}.
\label{cghz4}
\end{equation}
where $|+\rangle$ and $|-\rangle$ represent the positive and negative SU(2) oscillating states, $|1\rangle$ and $|2\rangle$ are the first and second round-trip locations of periodic orbits, here we change the polarization states from horizontal and vertical linear-polarization states $|H\rangle$ and $|V\rangle$ into diagonal and antidiagonal linear-polarization state $|D\rangle=\frac{1}{\sqrt{2}}(|H\rangle+|V\rangle)$ and $|A\rangle=\frac{1}{\sqrt{2}}(|H\rangle-|V\rangle)$ only because it is corresponding to our experimental description more conveniently.

\begin{figure*}
	\centering
	\includegraphics[width=\linewidth]{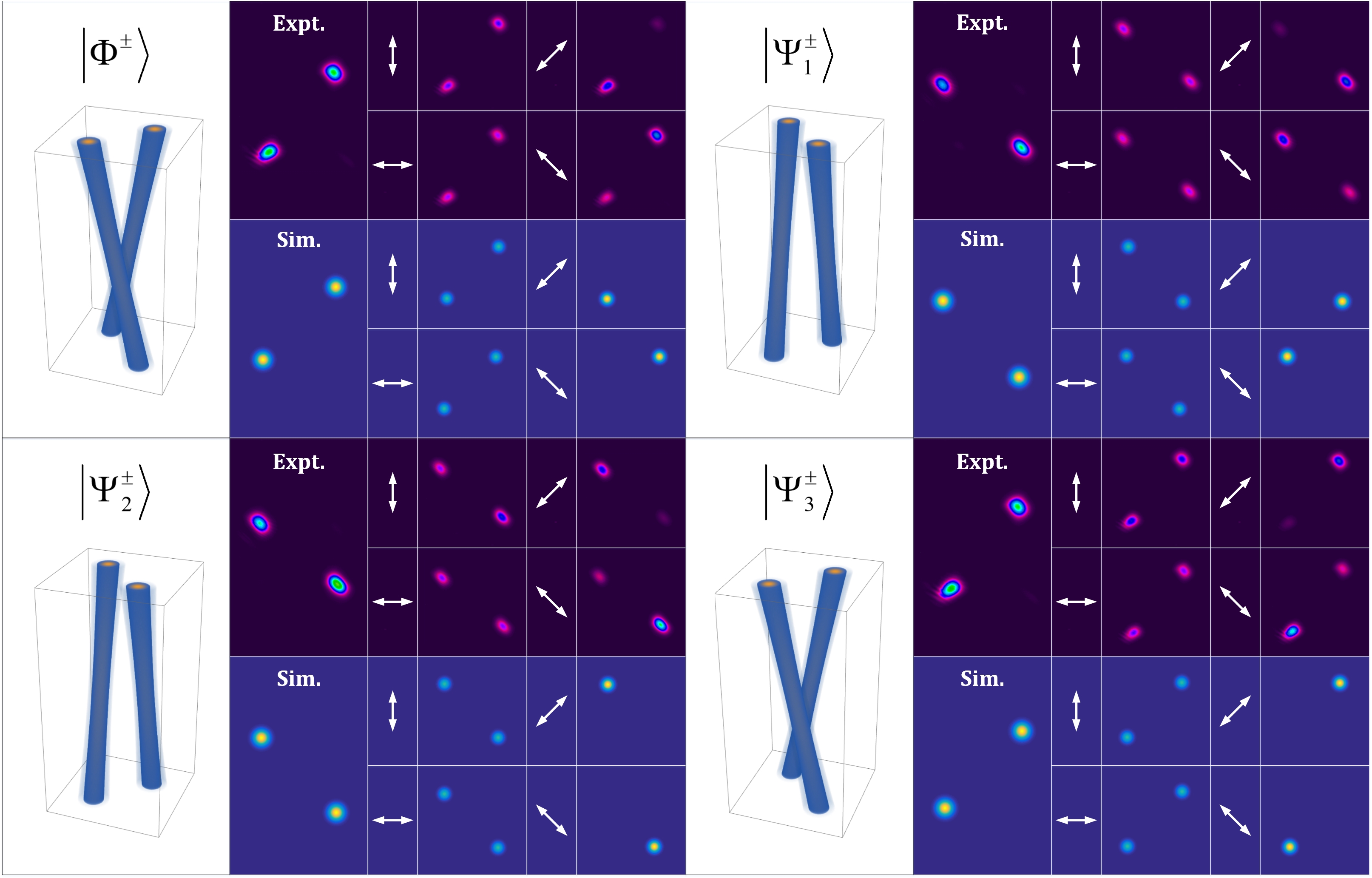}
	\caption{\footnotesize \textbf{Control of four maximumly entangled groups in GHZ states.} The experimental and theoretical results of four general SU(2) vector beams corresponding to the four maximumly entangled groups in classical GHZ states. The white arrow means the allowed polarization orientation of the polarizer. The plot on the left for each group shows the theoretical spatial wave-packet of the corresponding SU(2)-like vector beams. $|\Phi^\pm\rangle$: the SU(2)-like vector beam with diagonal intensity pattern where $|+1\rangle$ orbit with diagonal polarization $|-2\rangle$ orbit with anti-diagonal polarization; $|\Psi^\pm_1\rangle$: anti-diagonal intensity pattern, $|-1\rangle$ orbit with diagonal polarization $|+2\rangle$ orbit with anti-diagonal polarization; $|\Psi^\pm_2\rangle$: anti-diagonal intensity pattern, $|+2\rangle$ orbit with diagonal polarization $|-1\rangle$ orbit with anti-diagonal polarization; $|\Psi^\pm_3\rangle$: diagonal intensity pattern, $|-2\rangle$ orbit with diagonal polarization $|+1\rangle$ orbit with anti-diagonal polarization.} 
	\label{P}
\end{figure*}

\subsection{Projected Bell states from GHZ states}\label{AH}
Via intensity modulation by iris on I$_1$ or I$_2$ and SLM with ``$\pi/2|3\pi/2$'' or ``$3\pi/2|\pi/2$'' phase-step mask, four orthogonal vector beams corresponding to the four maximumly entangled groups $|\Phi^\pm\rangle$, $|\Psi^\pm_1\rangle$, $|\Psi^\pm_2\rangle$, and $|\Psi^\pm_3\rangle$ would be obtained. By using the polarizer, we measured the vector properties in these beams. Figure~\ref{P} shows the theoretical and experimental results of the intensity pattern of the four maximumly entangled groups before and after polarizer measure. For $|\Phi^\pm\rangle$ and $|\Psi^\pm_3\rangle$, the intensity pattern is diagonal, and polarizations are orthogonal at the corresponding orbit location, i.e. $|1\rangle|D\rangle$ with $|2\rangle|A\rangle$ and $|1\rangle|A\rangle$ with $|2\rangle|D\rangle$. For $|\Psi^\pm_1\rangle$ and $|\Psi^\pm_2\rangle$, the intensity pattern is antidiagonal, polarization distribution are $|1\rangle|D\rangle$ with $|2\rangle|A\rangle$ and $|1\rangle|A\rangle$ with $|2\rangle|D\rangle$. So far, we cannot say we obtain the vector beams completely expressed in 8-D space because we cannot distinguish the ``$\pm$'' signals and get the complete tomography of the 8 GHZ states.

Hereinafter, we demonstrate how to distinguish the ``$\pm$'' states in the four groups of maximum entanglement. Actually, the ``$+$'' and ``$-$'' states are corresponding to the classical orbits with phase differences of 0 and $\pi$. The phase difference of two classical light can be revealed by observation of interference fringes. In order to observe interference fringes of orbits, we should project the vector beams into a certain polarization for producing coherence. According to the properties of GHZ states, a GHZ state will be reduced into a Bell state after polarization projection. For classical GHZ states, different Bell states will also be obtained after polarization projection, yielded by: 
\begin{align}
\nonumber
\left| {{\Phi }^{\pm }} \right\rangle &= \frac{1}{\sqrt{2}}{\left( \left| + \right\rangle \left| 1 \right\rangle \frac{\left| H \right\rangle +\left| V \right\rangle }{\sqrt{2}}\pm \left| - \right\rangle \left| 2 \right\rangle \frac{\left| H \right\rangle -\left| V \right\rangle }{\sqrt{2}} \right)} \\ \nonumber
&=\frac{\left| + \right\rangle \left| 1 \right\rangle \pm \left| - \right\rangle \left| 2 \right\rangle }{2}\left| H \right\rangle +\frac{\left| + \right\rangle \left| 1 \right\rangle \mp \left| - \right\rangle \left| 2 \right\rangle }{2}\left| V \right\rangle \\ 
&=\frac{\left| {{\psi }^{\pm }} \right\rangle \left| H \right\rangle +\left| {{\psi }^{\mp }} \right\rangle \left| V \right\rangle }{\sqrt{2}},
\label{p} 
\end{align}
\begin{align}
\nonumber
\left| \Psi _{1}^{\pm } \right\rangle &=\frac{1}{\sqrt{2}}\left( \left| - \right\rangle \left| 1 \right\rangle \frac{\left| H \right\rangle +\left| V \right\rangle }{\sqrt{2}}\pm \left| + \right\rangle \left| 2 \right\rangle \frac{\left| H \right\rangle -\left| V \right\rangle }{\sqrt{2}} \right) \\ \nonumber
&=\frac{\left| - \right\rangle \left| 1 \right\rangle \pm \left| + \right\rangle \left| 2 \right\rangle }{2}\left| H \right\rangle +\frac{\left| - \right\rangle \left| 1 \right\rangle \mp \left| + \right\rangle \left| 2 \right\rangle }{2}\left| V \right\rangle \\ 
&=\frac{\pm \left| {{\phi }^{\pm }} \right\rangle \left| H \right\rangle \mp \left| {{\phi }^{\mp }} \right\rangle \left| V \right\rangle }{\sqrt{2}},
\label{p1} 
\end{align}
\begin{align}
\nonumber
\left| \Psi _{2}^{\pm } \right\rangle &= \frac{1}{\sqrt{2}}\left( \left| + \right\rangle \left| 2 \right\rangle \frac{\left| H \right\rangle +\left| V \right\rangle }{\sqrt{2}}\pm \left| - \right\rangle \left| 1 \right\rangle \frac{\left| H \right\rangle -\left| V \right\rangle }{\sqrt{2}} \right) \\ \nonumber
&=\frac{\left| + \right\rangle \left| 2 \right\rangle \pm \left| - \right\rangle \left| 1 \right\rangle }{2}\left| H \right\rangle +\frac{\left| + \right\rangle \left| 2 \right\rangle \mp \left| - \right\rangle \left| 1 \right\rangle }{2}\left| V \right\rangle \\ 
&=\frac{\left| {{\phi }^{\pm }} \right\rangle \left| H \right\rangle +\left| {{\phi }^{\mp }} \right\rangle \left| V \right\rangle }{\sqrt{2}},  
\label{p2} 
\end{align}
\begin{align}
\nonumber
\left| \Psi _{3}^{\pm } \right\rangle &= \frac{1}{\sqrt{2}}\left( \left| + \right\rangle \left| 1 \right\rangle \frac{\left| H \right\rangle -\left| V \right\rangle }{\sqrt{2}}\pm \left| - \right\rangle \left| 2 \right\rangle \frac{\left| H \right\rangle +\left| V \right\rangle }{\sqrt{2}} \right) \\ \nonumber
&=\frac{\left| + \right\rangle \left| 1 \right\rangle \pm \left| - \right\rangle \left| 2 \right\rangle }{2}\left| H \right\rangle -\frac{\left| + \right\rangle \left| 1 \right\rangle \mp \left| - \right\rangle \left| 2 \right\rangle }{2}\left| V \right\rangle \\ 
&=\frac{\left| {{\psi }^{\pm }} \right\rangle \left| H \right\rangle -\left| {{\psi }^{\mp }} \right\rangle \left|V \right\rangle }{\sqrt{2}},
\label{p3} 
\end{align}
where the classical Bell states should be:
\begin{equation}
\left| {{\phi }^{\pm }} \right\rangle =\frac{\left| + \right\rangle \left| 2 \right\rangle \pm \left| - \right\rangle \left| 1 \right\rangle }{\sqrt{2}},
\label{cb1}
\end{equation}
\begin{equation}
\left| {{\psi }^{\pm }} \right\rangle =\frac{\left| + \right\rangle \left| 1 \right\rangle \pm \left| - \right\rangle \left| 2 \right\rangle }{\sqrt{2}}.
\label{cb2}
\end{equation}

After projection onto $|H\rangle$ and $|V\rangle$ states, $|\Phi^\pm\rangle$ or $|\Psi^\pm_3\rangle$ state would be reduced to $|\psi^\pm\rangle$ and $|\psi^\mp\rangle$ states, $|\Psi^\pm_1\rangle$ or $|\Psi^\pm_2\rangle$ to $|\phi^\pm\rangle$ and $|\phi^\mp\rangle$. The ``$+$'' and ``$-$'' in Bell states can be distinguished by the complementary interferometric fringes of the corresponding phase difference of 0 and $\pi$ between two SU(2) orbits. For measuring $|\Phi^\pm\rangle$ and $|\Psi^\pm_3\rangle$, the CCD camera should be located at $z=-z_R$ position where $|-1\rangle$ and $|+2\rangle$ orbits are overlapped. For $|\Psi^\pm_1\rangle$ and $|\Psi^\pm_2\rangle$, the CCD camera should be located at $z=z_R$ position where $|+1\rangle$ and $|-2\rangle$ orbits are overlapped. Without polarization projection, the pattern shows no fringes because the light on the corresponding two orbits are incoherent. After projection on $|H\rangle$ or $|V\rangle$ states, different interference fringes would be observed for different reduced Bell states. For the group $|\Phi^\pm\rangle$, the ``$\pm$'' cannot be distinguished by the intensity patterns. However if we project the polarization onto $|H\rangle$ state to observe the pattern of $\langle H|\Phi^\pm\rangle$, pattern of original state $|\Phi^\pm\rangle$ will be reduced into Bell states $|\psi^\pm\rangle$ and two different patterns of complementary fringes will be observed, center-bright fringes for $|\psi^+\rangle$ (the original state should be $|\Phi^+\rangle$) and center-dark fringes for $|\psi^-\rangle$ (the original state should be $|\Phi^-\rangle$). We can also use projected state $\langle V|\Phi^\pm\rangle$ to distinguish the ``$\pm$'' that $\langle V|\Phi^+\rangle$ should be center-dark fringes corresponding to Bell state $|\psi^-\rangle$, and $\langle V|\Phi^-\rangle$ should be center-bright fringes corresponding to Bell state $|\psi^+\rangle$. In experiment, we can use a BK7 thin plate to cover one of the two orbits and rotate slightly to control the phase difference between them to control a phase difference of $\pi$, switching from ``$+$'' to ``$-$'' state. Other GHZ states can be generated by the similar way fulfilling a completed set in 8-D Hilbert space. 

\begin{figure*}
	\centering
	\includegraphics[width=\linewidth]{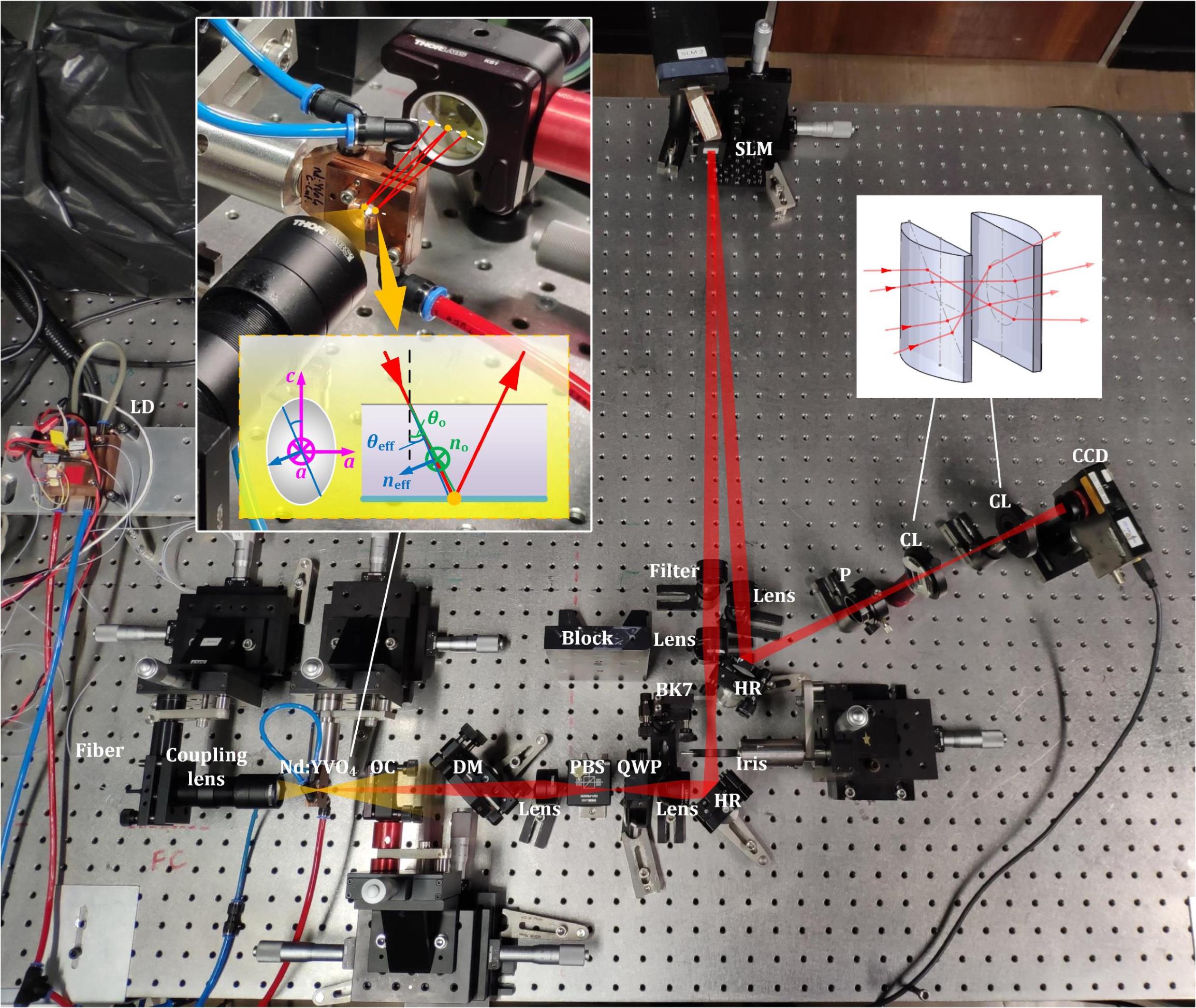}
	\caption{\footnotesize \textbf{Picture of experimental setup.} The actual picture is in accord with the schematic of experimental setup of Fig.~\ref{f3}, where one insert (left) shows the details of the arrangement of c-cut Nd:YVO$_4$, another insert (right) shows the two cylindrical lenses playing as astigmatic mode converter with the ray-represent of the transformation from a planar geometric mode to a vortex geometric mode. The trajectory of SU(2) geometric mode is depicted in the cavity. The index ellipsoide is depicted for determining the effective refraction index and angle of a geometric beam in crystal. OC: output coupler; DM: dichroic mirror; PBS: polarization splitting prism; QWP: quarter-wave plate; HR, high-reflective mirror; SLM, spatial light modulator; P: polarizer; CL: cylindrical lens; CCD: charge-coupled device.} 
	\label{expt}
\end{figure*}

\begin{figure*}
	\centering
	\includegraphics[width=0.62\linewidth]{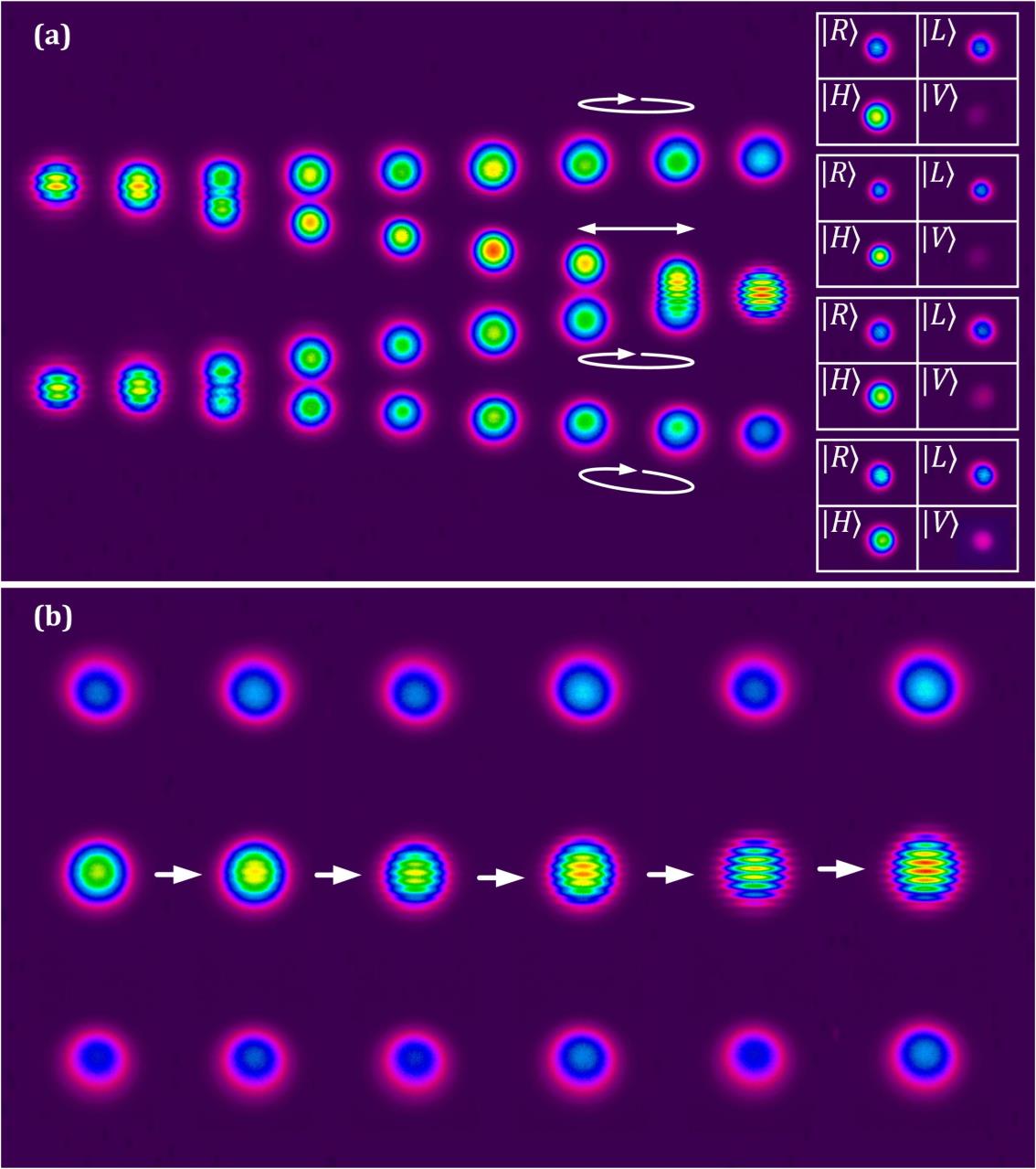}
	\caption{\footnotesize \textbf{Experimental generation of general SU(2) vector beams.} (a) The experimental measure of transverse patterns versus the propagation distance of an SU(2) geometric mode in state $|\Omega=1/4\rangle|\phi=\pi\rangle$, where different orbits show different intensities and polarizations.  (b) The experimentally observed evolution of the transverse pattern at $z=z_R$ plane with the adjustment of the pumping power and displacement.} 
	\label{gsu2}
\end{figure*}

\subsection{Generating general high-dimensional classical entanglement from a laser}\label{AI}
The general SU(2) vector beams with general high-dimensional entanglement can be directly generated in a cavity. In other words, a degenerate cavity with off-axis displacement should commonly generate the general SU(2) vector beams while normal SU(2) scalar beams are just specific cases. Actually, considering the anisotropism in gain medium induced by crystal cutting geometry and nonuniform thermal effect by asymmetric pumping, the geometric beams can undergo complex amplitude and polarization modulations in cavity and be output as vector fields~\cite{lu2012observation,lu2015generating}. In our experiment, we used a c-cut Nd:YVO$_4$ as gain medium to realize intracavity complex amplitude and polarization modulations. Nd:YVO$_4$ is a positive uniaxial crystal with anisotropic refractive indices and stimulated absorption and emission cross-sections~\cite{guo2019anisotropic}. The detailed schematic for arranging the crystal in our experiment is depicted in Fig.~\ref{expt}. In c-cut Nd:YVO$_4$, the principal c-axis is located on $z$-axis and other two a-axes on the $(x,y)$ transverse plane. For a light beam with normal incidence, there is no birefrigent effect; for a beam with a incident angle $\theta_\text{in}$, there is a birefrigent modulation where the vertical linear polarized component undergoes an ordinary refractive index $n_\text{o}$ and angle $\theta_\text{o}$ while the horizontal linear polarized component undergoes an effective refractive index $n_\text{eff}$ and angle $\theta_\text{eff}$ involved in the ordinary and extraordinary refractive indices $n_\text{o}$ and $n_\text{e}$:
\begin{equation}
n_\text{eff}=\frac{n_\text{o}n_\text{e}}{\sqrt{n_\text{e}^2\cos^2{\theta_\text{in}}+n_\text{o}^2\sin^2{\theta_\text{in}}}}.
\end{equation}
The difference of the two refractive indices leads to the phase retardation between the orthogonal polarized components, which can be represented as
\begin{equation}
\Delta=\frac{2\pi d}{\lambda_0}\left(\frac{n_\text{eff}}{\cos{\theta_\text{eff}}}-\frac{n_\text{o}}{\cos{\theta_\text{o}}}\right),\label{Del}
\end{equation}
where $d$ is the thickness of the crystal. According to Eq.~(\ref{Del}), when the laser crystal is given ($d$ and $\lambda_0$ is determined), the polarization control of geometric mode can be experimentally realized by two methods of: (1) modulating the refractive indices; (2) modulating the incidence angle, i.e. the included angle of classical orbits of geometric mode. The method-(1) can be realized by control of pump power, and the method-(2) can be realized by control of off-axis displacement.

\textbf{Modulation by pump power.} It has been proved that the temperature-dependent thermal effects in Nd:YVO$_4$ solid-state laser is highly related to the pump power, while the refractive indices, stimulated absorption and emission cross-sections are all related to the thermal effect~\cite{shen2017four,shen2017spatial,shen2018beam}. Thus the different powers corresponding to different thermal effects and then to different refractive indices. In SU(2) geometric mode, the pump spot is off-axis, thus nonuniform thermal effect is nonuniform and the polarization modulation is also nonuniform, resulting into the output of SU(2) vector beams.

\textbf{Modulation by off-axis displacement.} According to the ray-wave duality, the SU(2) geometric modes with different orders have different incident angles for various orbits and the ray trajectory is coupled with the pumping spot. Therefore, the control of pumping spot is related to the included angles of incidence rays in the classical trajectory. The different off-axis displacements corresponding to different transverse-orders of geometric mode result into different incidence angles. Note that the nonuniform thermal distribution and transverse-order comprehensively impact on the polarization of geometric mode.

Therefore, there should be asymmetrical complex amplitude and polarization modulations by general pumping control. For instance, Fig.~\ref{gsu2}(a) shows an experimental result of an SU(2) geometric mode in state $|\Omega=1/4\rangle|\phi=\pi\rangle$, where different orbits show different intensities and polarizations. For each orbit, the left- and right-handed circular polarized components are measured by a polarization grating, which uses geometric phase to diffract light into two beams in the $+1$ and $-1$ orders such that the two output beams have opposite circular polarizations, and the horizontal and vertical linear polarized components are measured by a rotating polarizer. Through the polarization components measurement, we can evaluate the actual polarizations of light on various orbits. The experimental results show that the SU(2) beam is a general vector state, i.e. the lights on different orbits have different polarization and intensity. 

The actual vector property can also be revealed by the interference fringes emerged at $z=0$ and $z=\pm z_R$ position. By adjusting the pumping power and displacement, we can modulate the vector field in the output general SU(2) vector beam, modulating the polarizations on $|+1\rangle$ and $|-2\rangle$ orbits from coherent to orthogonal. The evolution of the transverse pattern at $z=z_R$ shows the coherence between $|+1\rangle$ and $|-2\rangle$ orbits changing from weak to strong with the interference fringes from vague to clear, unraveling the polarizations on the two orbits changing into orthogonal states. This experimentally observed evolution is shown in Fig.~\ref{gsu2}(b).

In summary, using off-axis pumping in a frequency-degenerate cavity with c-cut Nd:YVO$_4$ as gain medium, general SU(2) vector beams with high-dimensional entanglement property can be generated. 

\subsection{Experimental setup details}\label{AJ}
The main setup is depicted in Fig.~\ref{f3} in the main text and Fig.~\ref{expt} for real picture. The frontend is a laser oscillator with cavity precisely adjusted into frequency-degenerate state $|\Omega=1/4\rangle$ and phase state $|\phi=\pi\rangle$. Through controlling the off-axis pumping, a ray-like SU(2) coherent state wave-packet can be generated and coupled with a W-shape classical geometric trajectory. The positive and negative oscillating states $|+\rangle$ and $|-\rangle$ share the overlapped trajectory. Here the signal is defined by the projection of photon propagating direction on $x$-axis. There are two round-trip orbits bouncing at flat mirror, i.e. the two V-shape orbits, noted as $|1\rangle$ and $|2\rangle$, for the coherent state $|\Omega=1/4\rangle|\phi=\pi\rangle$~\cite{rwd2}. The intracavity geometric trajectory is constituted by four periodic oscillating orbits $|+1\rangle$, $|+2\rangle$, $|-1\rangle$, and $|-2\rangle$, each of them contributes a ray-like orbit in the actual output SU(2) geometric mode. The planar geometric mode can be astigmatically converted into vortex geometric mode with OAM where the corresponding states $|\pm\rangle$, $|1\rangle$ and $|2\rangle$ are located at a hyperbolic ruled surface with orthogonal SU(2) coordinates as illustrated in the introduction of concept in the main text. 

A 808~nm fiber-coupled laser diode (LD) (FOCUSLIGHT, FL-FCSE08-7-808-200) was used as the pump source. With a telescope system with magnification about 1:1  constituted by two identical anti-reflective (AR) coated lenses (focal length $F=25$~mm), the pump light was focused into a c-cut Nd:YVO$_4$ slice-like crystal with dopant of 0.5~at.\% and thickness of 5~mm, which was wrapped in a copper heat sink and conductively water cooled at 18$^\circ$C. The outside surface of crystal was coated AR at 808~nm and high-reflective (HR) at 1064~nm and the inner surface AR at 1064~nm. A plano-concave mirror was used as the output coupler, where the radius of curvature is 100~mm and the transmittance is 10\% at 1064~nm for inner surface and AR for outside surface. 

The laser was firstly passed through a dichroic mirror (DM, 45$^\circ$ incidence, HR at pump light and AR at laser) for filtering residual pumping light. After lens focusing and passing through a polarization splitting prism (PBS) and quarter-wave plate (QWP), we can control the SU(2) beam as right-hand circular polarization state. Using a high-reflective mirror and lens, we can illuminate the SU(2) beam on a spatial light modulator (SLM) with beam waist location overlapped with the phase mask and incidence angle less than 5$^\circ$. A filter was used before the illumination on SLM for avoiding over-power damage. The phase modulation is only sensitive to horizontal linear polarization component, thus it can be used to modulate polarization and generate SU(2) structured vector beams by phase mask design. For a input light with right-hand circular polarization, if the phase mask plays a $\pi/2$ constant phase, the output will be diagonal linear polarization, and if $3\pi/2$, anti-diagonal linear polarization. Here we equally divided the phase mask into two parts, one for modulating the polarization on $|1\rangle$ and another for $|2\rangle$. When the SLM add divided phase ``$\pi/2|3\pi/2$'' on the mask, $|1\rangle|D\rangle$ and $|2\rangle|A\rangle$ states can be produced, and when ``$3\pi/2|\pi/2$'' on phase mask, $|1\rangle|A\rangle$ and $|2\rangle|D\rangle$ states are produced.

For introducing more structure control, we used an iris to make on-demand intensity modulation. Due to the special spatial structure of the SU(2) beam, different effects would be introduced for different location of iris application. When the iris with a proper aperture size was applied at negative Rayleigh length position (I$_1$ position), the orbits $|-1\rangle$ and $|+2\rangle$ could be blocked, resulting in a diagonal intensity pattern in the corresponding vortex SU(2) beam. When applied at positive Rayleigh length position (I$_2$ position), $|+1\rangle$ and $|-2\rangle$ could be blocked, resulting in anti-diagonal intensity pattern vortex SU(2) beam. A BK7 thin plate can be placed partially at the beam waist position for adding a phase difference for $|1\rangle$ or $|2\rangle$ states. After intensity modulation, the structured light can no longer be a normal SU(2) geometric mode, we here call the new structured light as general SU(2) beams and the beam after polarization modulation as general SU(2) vector beams. Because we used a c-cut Nd:YVO$_4$ providing anisotropic intracavity polarization modulation on geometric modes~\cite{guo2019anisotropic,lu2012observation}, our laser output should be a general SU(2) vector beam. Hereinafter, we will demonstrate that the general SU(2) vector beams can be measured completely by 3-DoF bases in 8-D Hilbert space and a complete set of GHZ states ($|\Phi^\pm\rangle$, $|\Psi^\pm_1\rangle$, $|\Psi^\pm_2\rangle$, and $|\Psi^\pm_3\rangle$) can be experimentally generated and controlled.

After the SLM modulation, the tomography measure system was designed. A astigmatic mode convertor (AMC) including two 45$^\circ$-inclined AR-coated cylindrical lenses was used to convert a planar general SU(2) beam into vortex beam with disjoint geometric orbits. After that, a charge coupled device (CCD) camera was used to identify the diagonal and anti-diagonal intensity patterns. A rotatable polarizer should be inset before the CCD to measure the vector properties and identify the diagonal and anti-diagonal linear polarization states for each orbit. The combination of this measure results in the tomography of for vector beams corresponding to the four maximumly entangled groups of $|\Phi^\pm\rangle$, $|\Psi^\pm_1\rangle$, $|\Psi^\pm_2\rangle$, and $|\Psi^\pm_3\rangle$. The ``$\pm$'' states are revealed by the phase difference on a certain orbit, which can be identified by the interference fringes of corresponding two orbits. For the SU(2) coherent state $|\Omega=1/4\rangle|\phi=\pi\rangle$, the interference of $|+\rangle$ and $|-\rangle$ orbits occurred at the beam waist, and the interference of $|1\rangle$ and $|2\rangle$ orbits occurred at the Rayleigh length $z_R$ away from the beam waist. For identifying ``$\pm$'' in GHZ states, the measure of the interference fringes is required. Before the measure of the interference fringes, a polarizer was used to project the polarization on a unified $|H\rangle$ or $|V\rangle$ state, in order to provide coherence, just corresponding to the process of Bell state projection. When the CCD was located at $z=-z_R$, the interference for $|-1\rangle$ and $|+2\rangle$ can be detected, and when $z=-z_R$, the interference for $|+1\rangle$ and $|-2\rangle$ can be detected.

\subsection{Towards higher-dimensional classically entangled state}\label{AK}
Hereto, we can already generate a complete set of GHZ states from a laser, sharing the same form of three-photon quantum entanglement. The general SU(2) modes also have potential to be used to generate even higher-dimensional entanglement states with the form of $N$-photon ($N>3$) quantum entanglement. To this end, we should find more DoFs to extend the dimension in classical entanglement such as OAM. As shown in Fig.~6(a-d) in the main text, two SU(2) modes with opposite OAM can be superposed together fulfilling a complete oscillating trajectory in a cavity, which can also be experimentally realized~\cite{lu2015generating}. After inhomogeneous intensity and polarization modulation, this kind of general SU(2) modes should be represented in 16-D Hilbert space like the 4-photon quantum entanglement:
\begin{align}
\nonumber
\left| \psi  \right\rangle=&\text{ }{{\alpha }_{1}}\left| +\ell  \right\rangle \left| + \right\rangle \left| 1 \right\rangle \left| H \right\rangle +{{\alpha }_{2}}\left| -\ell  \right\rangle \left| + \right\rangle \left| 1 \right\rangle \left| H \right\rangle \\ \nonumber
&+{{\alpha }_{3}}\left| +\ell  \right\rangle \left| + \right\rangle \left| 1 \right\rangle \left| V \right\rangle +{{\alpha }_{4}}\left| -\ell  \right\rangle \left| + \right\rangle \left| 1 \right\rangle \left| V \right\rangle \\ \nonumber
&+{{\alpha }_{5}}\left| +\ell  \right\rangle \left| - \right\rangle \left| 1 \right\rangle \left| H \right\rangle +{{\alpha }_{6}}\left| -\ell  \right\rangle \left| - \right\rangle \left| 1 \right\rangle \left| H \right\rangle\\ \nonumber
&+{{\alpha }_{7}}\left| +\ell  \right\rangle \left| - \right\rangle \left| 1 \right\rangle \left| V \right\rangle +{{\alpha }_{8}}\left| -\ell  \right\rangle \left| - \right\rangle \left| 1 \right\rangle \left| V \right\rangle\\ \nonumber
&+{{\alpha }_{9}}\left| +\ell  \right\rangle \left| + \right\rangle \left| 2 \right\rangle \left| H \right\rangle +{{\alpha }_{10}}\left| -\ell  \right\rangle \left| + \right\rangle \left| 2 \right\rangle \left| H \right\rangle\\ \nonumber
&+{{\alpha }_{11}}\left| +\ell  \right\rangle \left| + \right\rangle \left| 2 \right\rangle \left| V \right\rangle +{{\alpha }_{12}}\left| -\ell  \right\rangle \left| + \right\rangle \left| 2 \right\rangle \left| V \right\rangle\\ \nonumber
&+{{\alpha }_{13}}\left| +\ell  \right\rangle \left| - \right\rangle \left| 2 \right\rangle \left| H \right\rangle +{{\alpha }_{14}}\left| -\ell  \right\rangle \left| - \right\rangle \left| 2 \right\rangle \left| H \right\rangle\\
&+{{\alpha }_{15}}\left| +\ell  \right\rangle \left| - \right\rangle \left| 2 \right\rangle \left| V \right\rangle +{{\alpha }_{16}}\left| -\ell  \right\rangle \left| - \right\rangle \left| 2 \right\rangle \left| V \right\rangle,
\label{d16} 
\end{align}
with 16 (4 partite and 8 maximumly entangled group) GHZ states as eigenstates (just show the first maximumly entangled group here):
\begin{equation}
\left| {{\Phi }^{\pm }} \right\rangle =\frac{\left| +\ell \right\rangle\left| + \right\rangle \left| 1 \right\rangle \left| D \right\rangle \pm \left| -\ell \right\rangle\left| - \right\rangle \left| 2 \right\rangle \left| A \right\rangle }{\sqrt{2}}.
\label{ghzd16}
\end{equation}

For exploring more DoFs to extend the dimension in classical entanglement, we can utilize multi-LG SU(2) beams~\cite{lu2011generation,tuan2018characterization}, where LG beams replace the Gaussian beams along SU(2) orbits in a geometric mode based on ray-wave duality. Thus, there is a main OAM state $|+\ell\rangle$ along the propagation axis, but also OAM $|+m\rangle$ along an SU(2) orbit. The sub-OAM $|+m\rangle$ has potential to be a DoF to reach higher-dimensional entanglement. As shown in Fig.~6(e,f) in the main text, exotic beams can be obtained by the superposition of two multi-LG SU(2) beams with opposite main OAM, realizing the 32-D entangled stated with 5-partite 16-group GHZ states as eigenstates (just show the first maximumly entangled group here):
\begin{equation}
\left| {{\Phi }^{\pm }} \right\rangle =\frac{\left| +m \right\rangle\left| +\ell \right\rangle\left| + \right\rangle \left| 1 \right\rangle \left| D \right\rangle \pm \left| -m \right\rangle\left| -\ell \right\rangle\left| - \right\rangle \left| 2 \right\rangle \left| A \right\rangle }{\sqrt{2}}.
\label{ghzd32}
\end{equation}
 
Besides finding more DoFs, increasing the number of orbits can also extend the dimension. The above demonstrations are all at degenerate state $|\Omega=1/4\rangle$. If we can control higher-order degenerate state $|\Omega=P/Q\rangle$ ($P$ and $Q$ are co-prime integers, $Q$ is even), the general SU(2) modes would be extended into $4Q$-dimensional space:
\begin{equation}
\left|\psi\right\rangle=\sum\limits_{i=\pm \ell }{\sum\limits_{j=\pm}{\sum\limits_{k=1}^{{Q}/{2}}{\left| i \right\rangle \left| j \right\rangle \left| k \right\rangle \left( {{\alpha }_{ijk}}\left| H \right\rangle +{{\beta }_{ijk}}\left| V \right\rangle  \right)}}}.
\end{equation}
Involving the sub-OAM in multi-LG SU(2) beams, that can be further extended into $8Q$-dimensional space:
\begin{align}
\nonumber
\left|\psi\right\rangle=\sum\limits_{h=\pm m}\sum\limits_{i=\pm \ell }\sum\limits_{j=\pm}&\sum\limits_{k=1}^{{Q}/{2}}\left| h \right\rangle \left| i \right\rangle \left| j \right\rangle \left| k \right\rangle\\
&\times\left( {{\alpha }_{hijk}}\left| H \right\rangle +{{\beta }_{hijk}}\left| V \right\rangle  \right).
\end{align}
The realization of high-dimensional classical entanglement can pave the way for developing a myriad of novel applications of quantum mechanism using classical light.
}

\end{document}